\RequirePackage{fix-cm}
\documentclass[smallextended]{svjour3}       %
\smartqed  %
\usepackage{cite}
\usepackage{amsmath,amssymb,amsfonts}
\usepackage{graphicx}
\usepackage{textcomp}
\usepackage{booktabs} %
\usepackage{tikz}
\usepackage[inline]{enumitem}
\usetikzlibrary{arrows}
\usepackage{lipsum}
\usepackage{makecell}
\usepackage{fmtcount} %
\usepackage{array,multirow}
\usepackage{xspace}
\usepackage{url}
\usepackage{scalerel}
\usepackage{fancyvrb}
\usepackage{xcolor}
\usepackage{colortbl}
\usepackage{moreverb}
\usepackage{mdframed}
\usepackage{listings}

\def\postbreak{%
  \raisebox{0ex}[0ex][0ex]{\ensuremath{\hookrightarrow\space}}}

\lstdefinestyle{docker}{
	language=bash,
	morekeywords={RUN,FROM,MAINTAINER},
	numbers=left,
	showstringspaces=false,
	frame=none,
}
\surroundwithmdframed{lstlisting}

\usepackage{newfloat}
\DeclareFloatingEnvironment[
    fileext=los,
    listname={List of Listings},
    name=Listing,
    placement=!h,
    within=none,
]{customlisting}

\usepackage{alltt,xcolor}
\usepackage{balance}
\usepackage{tcolorbox}
\usepackage{blindtext}%
\tcbuselibrary{listings,breakable}
\tcbuselibrary{theorems}
\usepackage{framed}

\newtcbtheorem{summary}{Summary}{%
	theorem name,%
	colback=gray!5,%
	colframe=gray!35!gray,%
	fonttitle=\bfseries,title after break={Summary  -- \raggedleft Continued}%
}{summary}

\usepackage{framed}
\usepackage{soul}
\usepackage{amsmath,amssymb,amsfonts}
\usepackage{algorithmic}
\usepackage{graphicx}
\usepackage{textcomp}
\usepackage{graphicx}
\usepackage{tikz}
\usepackage{marvosym}
\usepackage{amsmath}
\usepackage[inline]{enumitem}
\usetikzlibrary{arrows}
\usepackage{lipsum}
\usepackage{balance}
\usepackage{makecell}
\usepackage{fmtcount} %
\usepackage{array,multirow}
\usepackage{xspace}
\usepackage{upgreek}
\usepackage{url}
\usepackage{hyperref}
\usepackage{amsmath,scalerel}

\usepackage{xcolor}
\definecolor{darkgreen}{rgb}{0,0.5,0.0}
\newcommand{\low}{{\color{darkgreen}low\xspace}}
\definecolor{darkblue}{rgb}{0.0, 0.0, 0.5}
\newcommand{\medium}{{\color{darkblue}medium\xspace}}  
\definecolor{orange}{rgb}{1.0, 0.5, 0.0}
\newcommand{\high}{{\color{orange}high\xspace}}
\definecolor{darkred}{rgb}{0.8,0.2,0.2}
\newcommand{\critical}{{\color{darkred}critical\xspace}}

\newcommand{\etal}{et al.\xspace}
\newcommand{\ie}{i.e.,\xspace}
\newcommand{\eg}{e.g.,\xspace}
\newcommand{\fig}[1]{Figure~\ref{#1}}
\newcommand{\tab}[1]{Table~\ref{#1}}

\newcommand{\sect}[1]{Section~\ref{#1}}

\newcommand{\cran}{{CRAN}\xspace}
\newcommand{\cpan}{{CPAN}\xspace}
\newcommand{\packagist}{{Packagist}\xspace}
\newcommand{\cargo}{{Cargo}\xspace}
\newcommand{\nuget}{{NuGet}\xspace}
\newcommand{\npm}{{npm}\xspace}
\newcommand{\pypi}{{PyPI}\xspace}
\newcommand{\rubygems}{{RubyGems}\xspace}

\newcommand{\github}{{GitHub}\xspace}

\newcommand{\java}{{Java}\xspace}
\newcommand{\javascript}{{JavaScript}\xspace}
\newcommand{\nodejs}{{Node.Js}\xspace}
\newcommand{\snyk}{{Snyk}\xspace}

\newcommand{\python}{{Python}\xspace}
\newcommand{\ruby}{{Ruby}\xspace}

\newcommand{\semver}{\emph{semver}\xspace}

\newcommand{\dockerhub}{\emph{Docker Hub}\xspace}

\newcommand{\librariesio}{\emph{libraries.io}\xspace}

\newcommand{\caret}{$^{\wedge}$}

\newcommand{\changed}[1]{{\color{black}#1}}
\newcommand{\minor}[1]{{\color{black}#1}}

\usepackage[disable]{todonotes}

\def\BibTeX{{\rm B\kern-.05em{\sc i\kern-.025em b}\kern-.08em
		T\kern-.1667em\lower.7ex\hbox{E}\kern-.125emX}}

\newcommand{\rqZero}{\changed{$RQ_0$: How prevalent are \minor{disclosed} vulnerabilities in \npm and \rubygems packages?}}
\newcommand{\rqOne}{\changed{$RQ_1$: How much time elapses until a vulnerability is disclosed?}}
\newcommand{\rqTwo}{\minor{$RQ_2$: For how long do packages remain affected by disclosed vulnerabilities?}}
\newcommand{\rqThree}{\changed{$RQ_3$: To \minor{what} extent are dependents exposed to their vulnerable dependencies?}}
\newcommand{\rqThreeA}{\changed{$RQ_3^a$: To \minor{what} extent are packages exposed to their vulnerable dependencies?}}
\newcommand{\rqThreeB}{\changed{$RQ_3^b$: To \minor{what} extent are external projects exposed to their vulnerable dependencies?}}
\newcommand{\rqFour}{\changed{$RQ_4$: How are vulnerabilities spread in the dependency tree?}}
\newcommand{\rqFourA}{\changed{$RQ_4^a$: How are vulnerabilities spread in the dependency trees of packages?}}
\newcommand{\rqFourB}{\changed{$RQ_4^b$: How are vulnerabilities spread in the dependency trees of external projects?}}
\newcommand{\rqFive}{$RQ_5$: Do exposed dependents upgrade their vulnerable dependencies when a vulnerability fix is released?}
\newcommand{\rqSix}{\minor{$RQ_6$: To what extent are dependents exposed to their vulnerable dependencies at their release time?}}
\newcommand{\rqSixA}{\minor{$RQ_6^a$: To what extent are packages exposed to their vulnerable dependencies at their release time?}}
\newcommand{\rqSixB}{\minor{$RQ_6^b$: To what extent are external projects exposed to their vulnerable dependencies at the date of their last commit?}}
\begin{document}
\title{On the Impact of Security Vulnerabilities\\ in the npm and RubyGems Dependency Networks}

\titlerunning{On the Impact of Security Vulnerabilities in Dependency Networks}
\author{Ahmed Zerouali \and
        Tom Mens \and
        Alexandre Decan \and
        Coen De Roover
}

\institute{A. Zerouali and C. De Roover\at
			  Vrije Universiteit Brussel,
			  Brussels, Belgium \\
              \email{ahmed.zerouali@vub.be, coen.de.roover@vub.be}  
              \and       %
              T. Mens and A. Decan \at
			Universit\'e de Mons, Mons, Belgium \\
			\email{tom.mens@umons.ac.be, alexandre.decan@umons.ac.be}  
}

\date{Received: date / Accepted: date}

\maketitle

\begin{abstract}
	The increasing interest in open source software has led to the emergence of large language-specific package distributions of reusable software libraries, such as \npm and \rubygems. These software packages can be subject to vulnerabilities that may expose dependent packages through explicitly declared dependencies.
	\minor{Using Snyk's vulnerability database,} this article empirically studies vulnerabilities affecting \npm and \rubygems packages. We analyse how and when these vulnerabilities are disclosed and fixed, and how their prevalence changes over time. We also analyse how vulnerable packages expose their direct and indirect dependents to vulnerabilities. We distinguish between two types of dependents: packages distributed via the package manager, and external \github projects depending on npm packages.
	We observe that the number of vulnerabilities in \npm is increasing and being disclosed faster than vulnerabilities in \rubygems.
	For both package distributions, the time required to \changed{disclose} vulnerabilities is increasing over time. 
	\changed{Vulnerabilities in \npm packages affect a median of 30 package releases, while this is 59 releases in \rubygems packages.}
	 A large proportion of external \github projects is exposed to vulnerabilities coming from direct or indirect dependencies.
	\changed{33\% and 40\% of dependency vulnerabilities to which projects and packages are exposed, respectively, have their fixes in more recent releases within the same major release range of the used dependency. Our findings reveal that more effort is needed to better secure open source package distributions.
	}

\end{abstract}

\textbf{Keywords:} security vulnerabilities, \npm, \rubygems, vulnerable packages

\section{Introduction}
\label{sec:intro}

Due to the increasing popularity and use of open source software (OSS), a large number of OSS ecosystems have emerged, 
containing huge collections of interdependent software packages. \changed{These ecosystems are usually supported by large communities of contributors, and can be considered as software supply chains formed by upstream transitive dependencies of packages and their downstream dependents.}
Typical examples of such ecosystems are package distributions for specific programming languages (\eg~\npm for \javascript, \pypi for \python, \rubygems for \ruby), totalling millions of interdependent reusable libraries maintained by hundreds of thousands of developers, and used by millions of software projects in a daily basis.
\changed{Given the sheer size of package distributions, combined with their open source nature, \minor{many packages are being affected by known or unknown vulnerabilities.}
Since package distributions are known to form huge and tightly interconnected dependency networks~\cite{Decan2019}, a single vulnerable package may potentially expose a considerable fraction of the package dependency network to its vulnerabilities~\cite{decan2018impact}.
This exposure does not even stop at the boundaries of the package distribution, since dependent external software projects may also become exposed to these vulnerabilities~\cite{Lauinger2017}.}
According to a study carried out by \snyk~\cite{snyk2017}, one of the leading companies in analysing and detecting software vulnerabilities in \nodejs and \ruby packages, 77\% of the 430k websites run at least one front-end package with a known security vulnerability in place.

\changed{This exposure of software to vulnerabilities in its dependencies} is considered as one of the OWASP top 10 application security risks~\cite{topOWASP}.
There are many examples of such cases.
For example, in November 2018 the widely used \npm package \texttt{event-stream} was found to depend on a malicious package named \texttt{flatmap-stream}~\footnote{\url{https://github.com/dominictarr/event-stream/issues/116}} containing a Bitcoin-siphoning malware. The \texttt{event-stream} package is very popular \changed{getting} roughly two million downloads per week. Just because it was used as a dependency in \texttt{event-stream}, the malicious \texttt{flatmap-stream} package was downloaded millions of times since its inclusion in September 2018 and until its discovery and removal.
The main objective of this paper is \changed{therefore to empirically analyse and quantify the impact of vulnerabilities in open source packages, on transitively dependent packages that are shared through the same package distribution as well as on external projects distributed via \github that rely on such packages.} \changed{We conduct our study on two different package distributions, \npm and \rubygems. All along our research questions, we compare between the results found in these two package distributions. More specifically, we answer the following research questions:}
\begin{itemize}
	\item \textbf{\rqZero}
	\changed{This preliminary research question explores the dataset of vulnerabilities extracted from \snyk's database and provides insights about their characteristics and evolution over time. }
	\item \textbf{\rqOne}
	\changed{By answering this question for both \npm and \rubygems, security researchers in these ecosystems will be able to assess how quick they are in finding and disclosing vulnerabilities. Users of these two ecosystems will be able to know which community has active security researchers which will eventually help them to assess their trust on the third-party packages they depend on.}
	\item \textbf{\rqTwo}
	\changed{Package dependents will gain insights about the number of dependency releases they should expect to be affected by a newly disclosed vulnerability and how many more releases are going to be affected by the same vulnerability even after its disclosure. The answer to this question will also help dependents to know to which type of package releases (\ie major, minor or patch) they should update to have their dependency vulnerabilities fixed.}
	\item \textbf{\rqThree}
	\changed{Vulnerable dependencies can expose their dependents to vulnerable code that might lead to security breaches. We will identify all direct and indirect vulnerable dependencies that are exposing packages and external \github projects, and characterize their vulnerabilities. \minor{In this and the next two research questions we study dependents as if they were deployed on 12 January 2020 (\ie the snapshot of the vulnerability dataset).}}
	\item \textbf{\rqFour}
	\changed{The answer to this question will inform us how deep in the dependency tree we can find vulnerabilities to which packages and external \github projects are exposed. This helps us to quantify the transitive impact that vulnerable packages may have on their transitive dependents.}
	\item \textbf{\rqFive}
	\changed{We will quantify how much dependent packages and dependent external \github projects would benefit from updating their dependencies for which there is a known fix available. We will also report on the number of vulnerable dependencies that could be reduced by only doing backward compatible updates.}
	\item \textbf{\rqSix}
	\minor{We will identify the dependencies that were only affected by vulnerabilities disclosed before each dependent's release time. Answering this research question will help us to assess whether developers are careful about incorporating dependencies with already disclosed vulnerabilities.}
\end{itemize}

The remainder of this article is structured as follows: \sect{sec:related} discusses related work and highlights the differences to previous studies. \sect{sec:method} explains the research method and the data extraction process, and presents a preliminary analysis of the selected dataset. \sect{sec:results} empirically studies the research questions for the \npm and \rubygems package distributions. \sect{sec:discussion} highlights the novel contributions, discusses our findings, and outlines possible directions for future work.
\sect{sec:threats} discusses the threats to the validity of this work and \sect{sec:conclusion} concludes.\section{Related work}
\label{sec:related}
\subsection{\changed{Terminology}}
\label{subsec:terminology}
This section introduces the terminology used throughout this article. All main terms are highlighted in \textbf{boldface}.

\textbf{Package distributions} (such as \npm and \rubygems) are collections of (typically open source) software \textbf{packages}, distributed through some package registry.
Each of these packages has one or more \textbf{releases}.
New releases of a package are called \textbf{package updates}.

Each package release is denoted by a unique \textbf{version number}. The version number reflects the sequential order of all releases of a package.
\textbf{Semantic versioning}, hereafter abbreviated as \semver\footnote{See \url{https://semver.org}}, proposes a multi-component version numbering scheme \textsf{major.minor.patch[-tag]}
to specify the type of changes that have been made in a package update.
Backward incompatible changes require an increment of the \textbf{major} version component, important backward compatible changes (\eg adding new functionality that does not affect existing dependents) require an increment of the \textbf{minor} component, and backward compatible bug fixes require an increment of the \textbf{patch} component.

The main purpose of package distributions is to facilitate software reuse. To do so,
a package release $R$ can explicitly declare a \textbf{dependency relation} to another package $P$.
$R$ will be called a \textbf{(direct) dependent} of $P$, while $P$ will be called a \textbf{(direct) dependency} of $R$.
Dependency relations come with {\bf dependency constraints}
that specify which releases of the dependency $P$
are allowed to be selected for installation when $R$ is installed. Such constraints express a \textbf{version range}. For example, constraint \verb|<2.0.0| defines the version range \verb|[0.0.0, 2.0.0)|, signifying that \emph{any} release below version 2.0.0 of the dependency is allowed to be installed. The highest available version within this range will be selected for installation by the package manager.
Combining \semver with dependency constraints enables maintainers of dependents to restrict the version range of dependencies to those releases that are expected to be backward compatible~\cite{decan2019package}.
For exemple, a dependency relation in an \npm release $R$ could express a constraint \verb|^1.2.3| to allow the version range \verb|[1.2.3,2.0.0)| of backward compatible releases. In \rubygems, dependency constraint \verb|~>1.2| would allow the version range \verb|[1.2.0,2.0.0)| of backward compatible releases.
The collection of all package releases and their dependencies in a package distribution forms a \textbf{package dependency network}.
A release $R$ is an \textbf{indirect dependent} of another release $D$ if there is a chain of length 2 or longer between them in the dependency network. Conversely $D$ will be called an \textbf{indirect dependency} of $R$.
We will refer to the union of direct and indirect dependents (respectively, dependencies) of a release as \textbf{transitive dependents} (respectively, dependencies) of that release.

\changed{In the context of this paper and more specifically in $RQ_3$ to $RQ_6$, we only study the vulnerabilities coming from dependencies used in the latest release of each package in \npm and \rubygems. \minor{Because of this, we will occasionally use the term package to refer to the latest release of a dependent package.} 
By abuse of terminology we declare a \emph{package} to be a (direct/indirect) dependent of $P$ if its latest available \emph{release} depends (directly or indirectly) on $P$.}

Not only \emph{packages} can depend on other packages within a package dependency network, but the same is true for \emph{external projects} that are developed and/or distributed outside of the package distribution (\eg on \github). By extension of the term dependent package, we use the term \textbf{dependent project} to refer to a \github repository containing an external software project that (directly or indirectly) depends on one of the packages of the considered package distribution. For example, the project \textsf{Atom}~\footnote{\url{https://github.com/atom/atom/blob/master/package.json}} is a dependent of \npm package \textsf{mocha}~\footnote{\url{https://www.npmjs.com/package/mocha}}. Similarly, \textsf{Discourse}~\footnote{\url{https://github.com/discourse/discourse/blob/master/Gemfile}} is a dependent project of \rubygems package \textsf{json}~\footnote{\url{https://rubygems.org/gems/json}}.
A \textbf{vulnerability} is a known reported security threat that affects some releases of some packages in the package distribution. The packages will be called \textbf{vulnerable packages} and their affected releases will be called \textbf{vulnerable releases}.
The package's vulnerability is \textbf{fixed} as soon as a package update that is no longer affected by the vulnerability becomes available.
A \textbf{vulnerable dependency} is a vulnerable release that is used as a dependency by another package or project (directly or indirectly).
Vulnerable dependencies can expose their dependents to the vulnerability.
We refer to those as \textbf{exposed dependents}. We can distinguish between \emph{directly exposed} dependents (if a direct dependency is vulnerable) and \emph{indirectly exposed} dependents (if an indirect dependency is vulnerable).
To distinguish between dependent releases and dependent projects that may be exposed, we use the terms \textbf{exposed package} (release) and \textbf{exposed project}, respectively.

\subsection{\changed{Package dependency networks}}
\label{subsec:pdn}
Software dependency management and package dependency networks have been subject to many research studies for different software ecosystems. Wittern \etal\cite{wittern2016look} examined the npm ecosystem in an extensive study that covers package descriptions, the dependencies among them, download metrics, and the use of \npm packages in publicly available repositories. 
One of their findings is that the number of \npm packages and their updates is growing superlinearly. They also observed that packages are increasingly connected through dependencies. More than 80\% of npm packages have at least one direct dependency. %
Kikas \etal\cite{kikas2017structure} analysed the dependency network structure and evolution of the JavaScript, Ruby, and Rust package distributions. One of their findings is that the number of transitive
dependencies for JavaScript has grown by 60\% in 2016. They also found that the negative consequences of removing a popular package (\eg the left-pad incident~\footnote{https://github.com/left-pad/left-pad/issues/4}) are increasing.
In a more extensive study, Decan \etal\cite{decan2017empirical} empirically compared the impact of dependency issues in the \npm, \cran and \rubygems package distributions. A follow-up study~\cite{Decan2019} expanded this comparison with four more distributions, namely \cpan, \packagist, \cargo and \nuget. They observed important differences between the considered ecosystems that are related to ecosystem specific factors. Similarly, Bogart \etal~\cite{bogart2016break} performed multiple case studies of three software ecosystems with different tooling and philosophies toward change (Eclipse, \cran, and \npm), to understand how developers make decisions about change and change-related costs and what practices, tooling, and policies are used. %
They found that the three ecosystems differ significantly in their practices, policies and community values. \changed{Gonzalez-Barahona \etal~\cite{gonzalez2017technical,zerouali2018empirical,zerouali2021multi,decan2018evolution} introduced the notion of technical lag to quantify the degree of outdatedness of packages and dependencies, along different dimensions, including time lag, version lag and vulnerability lag. }

\subsection{\changed{Security vulnerabilities}}
\label{subsec:sv}
\minor{Software vulnerabilities are discovered on a daily basis, and need to be identified and fixed as soon as possible.}
This explains the many studies conducted by software engineering researchers on the matter (\eg~\cite{pham_detecting_2010,Shin2010TSE,Pashchenko2018,ruohonen2018empirical,chinthanet2020code}). Several researchers observed that outdated dependencies are a potential source of security vulnerabilities.

Cox~\etal\cite{cox2015measuring} analysed 75 Java projects that manage their dependencies through Maven. They observed that projects using outdated dependencies were four times more likely to have security issues and backward incompatibilities than systems that were up-to-date. Gkortzis \etal~\cite{Gkortzis2020jss} studied the relationship between software reuse and security threats by empirically investigating 1,244 open-source Java projects to explore the distribution of vulnerabilities between the code created by developers and the code reused through dependencies. Based on a static analysis of the source code, they observed that large projects are associated with a higher number of potential vulnerabilities. Additionally, they found that the number of dependencies in a project is strongly correlated to its number of vulnerabilities. \changed{Massacci \etal~\cite{massacci2021technical} investigated whether leveraging on FOSS Maven-based Java libraries is a potential source of security vulnerabilities. They found that small and medium libraries have disproportionately more leverage on FOSS dependencies in comparison to large libraries. They also found that libraries with higher leverage have 1.6 higher odds of being vulnerable in comparison to the libraries with lower leverage. }

Ponta~\etal\cite{Ponta2020EMSE} presented a code-centric and usage-based approach to detecting and assessing OSS vulnerabilities, and to determining their reachability through direct and transitive dependencies of \java applications. Their combination of static and dynamic analysis improves upon the state of the art which commonly considers dependency meta-data only without verifying to which  dependents the vulnerability actually propagates. The Eclipse Steady tool instantiates the approach, and has been shown to report fewer false positives (\ie vulnerabilities that do not really propagate to dependencies) than earlier tools.
In a similar vein, Zapata \etal\cite{zapata2018towards} carried out an empirical study that analysed vulnerable dependency migrations at the function level for 60 JavaScript packages. They provided evidence that many outdated projects are free of vulnerabilities as they do not really rely on the functionality affected by the vulnerability. Because of this, the authors claim that security vulnerability analysis at package dependency level is likely to be an overestimation. 

Decan \etal~\cite{decan2018impact} conducted an empirical analysis of 399 vulnerabilities reported in the \npm package dependency network containing over 610k \javascript packages in 2017.
They analysed how and when these vulnerabilities are \changed{disclosed} and to which extent this affects directly dependent packages.
They did not consider the effect of vulnerabilities on transitive dependents, nor did they study the impact on external \github projects depending on \npm packages.
They observed that it often takes a long time before an introduced vulnerability is \changed{disclosed}.
A non-negligible proportion of vulnerabilities (15\%) are considered to be more risky because they are fixed only after public announcement of the vulnerability, or not fixed at all. 
They also found that the presence of package dependency constraints plays an important role in vulnerabilities not being fixed, mainly because the imposed constraints prevent fixes from being installed.

Zimmermann \etal\cite{zimmermann2019small} studied 609 vulnerabilities in \npm packages, providing evidence that this ecosystem suffers from single points of failure, \ie a very small number of maintainer accounts could be used to inject malicious code into the majority of all packages (\eg the \textsf{event-stream} incident~\footnote{\url{https://www.theregister.com/2018/11/26/npm_repo_bitcoin_stealer/}}). This problem increases with time, and unmaintained packages threaten large code bases, as the lack of maintenance causes many packages to depend on vulnerable code, even years after a vulnerability has become public. They studied the transitive impact of vulnerable dependencies, as well as the problems related to lack of maintenance of vulnerable packages. 

Alfadel~\etal\cite{alfadelempirical} carried out a study on 550 vulnerability reports affecting 252 \python packages from \pypi. They found that the number of vulnerabilities \changed{disclosed} in \python packages increases over time, and some take more than 3 years to be \changed{disclosed}. \changed{Meneely~\etal~\cite{meneely2013patch} inspected 68 vulnerabilities in the Apache HTTP server and traced them back to the first commits that contributed to the vulnerable code. They manually found 124 Vulnerability-Contributing Commits (VCCs). After analyzing these VCCs, they found that VCCs have more than twice as much code churn on average than non-VCCs. They also observed that commits authored by new developers have more chances to be VCCs than other commits.}

Prana \etal~\cite{Prana2021EMSE} analysed vulnerabilities in open-source libraries used by 450 software projects written in \java, \python, and \ruby. Using an industrial software composition analysis tool, they scanned versions of the sample projects after each commit. They found that project activity level, popularity, and developer experience do not translate into better or worse handling of dependency vulnerabilities. As a recommendation to software developers, they highlighted the importance of managing the number of dependencies and of performing timely updates.

\changed{Pashchenko \etal~\cite{Pashchenko2018} studied the over-inflation problem of academic and industrial approaches for reporting vulnerable dependencies in OSS software. After inspecting 200 Java libraries they found that 20\% of their dependencies affected by a known vulnerability are not deployed, and therefore, they do not represent a danger to the analyzed library because they cannot be exploited in practice. They also found that 81\% of vulnerable direct dependencies may be fixed by simply updating to a new version. In our article, we follow the procedure recommended by Pashchenko \etal~\cite{Pashchenko2018} by only focusing on run-time dependencies which are essential for deployment.}
\subsection{\changed{Novelty of our contribution}}
\label{subsec:noc}
The empirical study proposed in this article expands upon previous work in different ways.
We conduct a quantitative comparison of vulnerabilities in both the \npm and \rubygems package dependency networks based on a more recent dataset of packages and their vulnerabilities (2020). We study the impact of a large set of 2,786 vulnerabilities, of which 2,118 for \npm and 668 for \rubygems while grouping them by their severity levels. We also consider dependencies of external \github projects on vulnerable \npm and \rubygems packages. 
For the latter, we are the first to study the prevalence, \changed{disclosure} and fixing time of their vulnerabilities.
When studying the impact of vulnerable packages on their dependents, we do not only focus on direct dependencies, but also consider the indirect ones. For those indirect dependencies we study the evolution and \changed{spread} of vulnerable indirect dependencies at different levels in the dependency tree. Finally, we are the first to compare the vulnerability of packages distributed via package distributions with the vulnerability of \github projects that use these packages. \changed{Such a comparison is important since packages are supposed to be reused as libraries, their maintainers are supposed to be more careful than developers of external projects that just depend on these reusable libraries and that are much less likely to have other projects depending on them.} We are also the first to report results on how vulnerability \changed{disclosure} time duration is evolving over time.
\section{Dataset}
\label{sec:method}

This paper analyzes \changed{the Common Vulnerabilities and Exposures (CVE\footnote{\url{cve.mitre.org}}) affecting the \npm and \rubygems distributions of reusable software packages.
Both package distributions are well-established and mature (\rubygems was created in 2004 and \npm in 2010) and both recommend \semver for package releases~\cite{decan2019package}. Due to an important difference in popularity of the targeted programming language (JavaScript and Ruby, respectively), the number of packages distributed through \npm is an order of magnitude higher than the number of packages distributed through \rubygems.
We focus on these package distributions because they have a community that is actively looking for and reporting vulnerabilities. The vulnerabilities in both distributions are reported and tracked by well-known Central Numbering Authorities (CNA)}.

\subsection{Vulnerability dataset}
\label{subsec:vulnerability}
To detect vulnerabilities in \npm and \rubygems packages we rely on a database of vulnerability reports of third-party package vulnerabilities collected by the continuous security monitoring service \snyk~\footnote{\url{https://snyk.io/vuln}}.
We received a snapshot of this vulnerability database on 17 April 2020.
 \changed{This snapshot contained 2,874 vulnerability reports for the considered package distributions, of which 2,188 for \npm and 686 for \rubygems. The higher number of reported vulnerabilities for \npm can be explained by the higher number of packages contained in it.}
 
 Each vulnerability report contains information about the affected package, the range of affected releases, the severity of the vulnerability \minor{as reported by Snyk's security team}, its origin (\ie the package distribution), the date of \changed{disclosure}, the date when it was published in the database, the first fixed version (if available), and the unique CVE identifier. 
\fig{fig:example-vul} shows an example of a vulnerability report of the popular \rubygems package \texttt{rest-client}~\footnote{\url{https://snyk.io/vuln/SNYK-RUBY-RESTCLIENT-459900}}. Vulnerability reports for \npm packages contain similar information.

\begin{figure}[!ht]
	\centering
	
	\begin{tabular}{|ll|}
		\hline
		\textbf{Vulnerability name:}                      & Malicious Package                           \\
		\textbf{Severity:}                   & critical                                    \\
		\textbf{Affected package:}           & rest-client                                 \\
		\textbf{Affected versions:}          & \textgreater{}=1.6.10, \textless{}1.7.0.rc1 \\
		\textbf{Package manager:}            & RubyGems                                    \\
		\textbf{\changed{Disclosure} date:}             & 2019-08-19                                  \\
		\textbf{Publication date:}           & 2019-08-20                                  \\
		\textbf{Version with the first fix:} & 1.7.0.rc1                                   \\
		\textbf{CVE identifier:} & CVE-2019-15224 \\ \hline
	\end{tabular}
	\caption{Excerpt of vulnerability report for \rubygems package {\sf rest-client}.}
	\label{fig:example-vul}
\end{figure}

\subsection{Dependency dataset of packages and external projects}
\label{subsec:dependency}
Using %
version 1.6.0 of the \librariesio Open Source Repository and Dependency Dataset~\cite{librariesio2020Jan} that was released on 12 January 2020, we identified all \emph{package releases} in the \npm and \rubygems package distributions. \changed{As this dataset was released three months before the snapshot date of the \snyk vulnerability database (17 April 2020), it does not contain releases that are affected by vulnerabilities disclosed after 12 January 2020. We therefore ignore these releases in our analysis.%

We also identified all \emph{external projects} hosted on GitHub and referenced in \librariesio dataset as depending on \npm or \rubygems packages. Repositories of these projects do not correspond to the development history of any of the considered packages, and are not forked from any other repository. This way we ensure that packages and external projects are mutually exclusive, and we avoid considering the same projects multiple times in the analysis.}

A manual analysis revealed that \librariesio is not always accurate about the dependencies used by external projects. \changed{More specifically, we occasionally observed a mix between run-time and development dependencies for projects that use \npm, \ie dependencies that are declared as development dependencies in the project's repository in \github are reported as run-time dependencies in \librariesio.
Because of this inaccuracy,} we only relied on \librariesio 
to identify the \changed{names} of the most starred projects, \ie those who received 90\% of all stars within the same ecosystem (\npm or \rubygems). \changed{Overall, the selected \github projects received 5.37M stars out of 5.96M. The minimum number of stars found in the resulting dataset was 62. Afterwards,} to determine the \npm and \rubygems package dependencies for these projects, we extracted and analysed their \emph{package.json} and \emph{Gemfile} from \github, which are the files in which the dependency metadata is stored for the respective package distributions.

As older package releases are more likely to be exposed to vulnerable packages than recent versions~\cite{cox2015measuring,zerouali2019saner}, this might bias the analysis results.
We therefore decided to focus only on the latest available version of each considered package, and on the snapshot of the last commit before 12 January 2020 of each considered external \github project.
We also decided to focus only on packages and external projects with {\em run-time dependencies}, thereby ignoring development and optional dependencies. Run-time dependencies are needed to deploy and run the dependent in production, while development dependencies are only needed  while the dependent is being developed (\eg for testing dependencies). We ignore the latter in our study because they are unlikely to affect the production environment\changed{~\cite{Pashchenko2018}}.

\begin{figure}[!ht]
	\begin{center}
		\setlength{\unitlength}{1pt}
		\footnotesize
		\includegraphics[width=0.98\columnwidth]{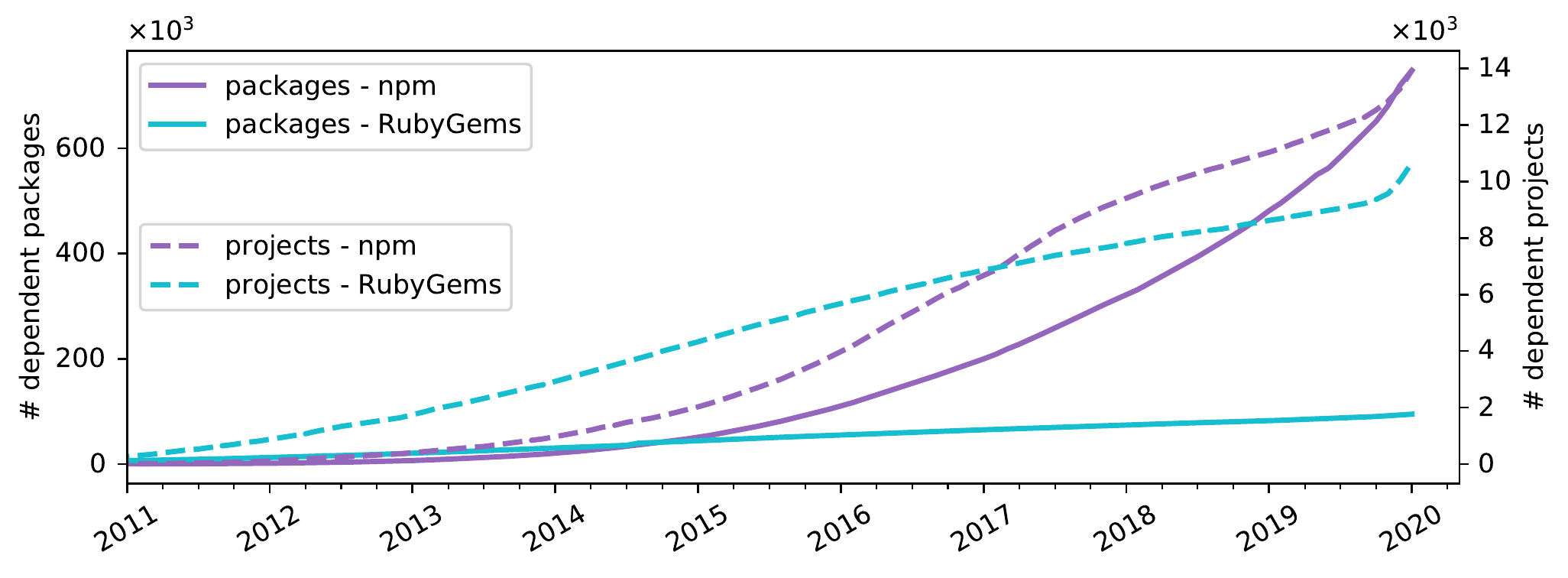}
		\caption{Evolution of the cumulative number of packages (straight lines, using the scale on the left y-axis) and external \github projects (dotted lines, using the scale on the right y-axis) for \npm and \rubygems}
		\label{fig:packages_evolution}
	\end{center}
\end{figure}

Based on all of the above, we selected the latest available releases in \librariesio of 842,697 packages \changed{(748,026 for \npm and 94,671 for \rubygems) and 24,593 external projects hosted on \github (13,930 using \npm packages and 10,663 using \rubygems packages)} with run-time dependencies. %
\fig{fig:packages_evolution} shows the evolution over time of the cumulative number of packages (left y-axis) and external projects (right y-axis) considered in this study, grouped by package distribution.

Using the dependency constraint resolver proposed in~\cite{decan2019package}, which supports several package distributions, we determined the appropriate version of packages to be installed for each dependent according to the constraints for its run-time dependencies. 
As some constraints may resolve to different versions at different points in time, we use the \librariesio snapshot date as the resolution date. This implies that we study vulnerabilities in packages and external projects as if they were installed or deployed on 12 January 2020.

\begin{figure}[!ht]
	\begin{center}
		\setlength{\unitlength}{1pt}
		\footnotesize
		\includegraphics[width=0.98\columnwidth]{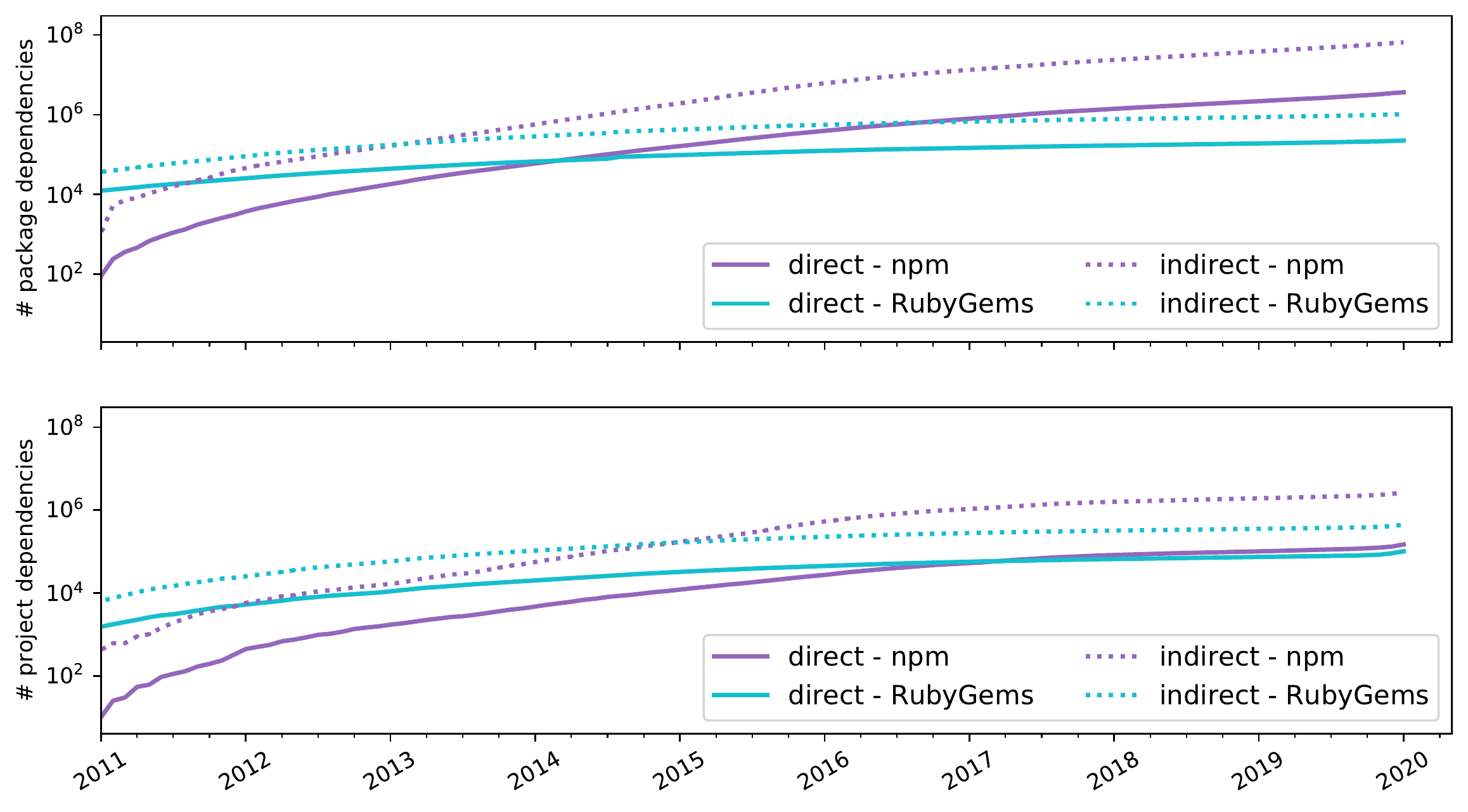}
		\caption{Monthly evolution of the cumulative number of direct and indirect dependencies for packages (top figure) and external projects (bottom figure) for \npm and \rubygems.}
		\label{fig:dependency_evolution}
	\end{center}
\end{figure}

Having determined the versions of all direct dependencies, we turn to the indirect ones. All considered \npm packages have a total of 68,597,413 dependencies of which 3,638,361 are direct (\ie 5.3\%), while all \rubygems packages have 1,258,829 dependencies of which 224,959 are direct (\ie 17.9\%).
The considered external projects have 2,814,544 \npm dependencies of which 147,622 are direct (\ie 5.2\%), and 544,506 \rubygems dependencies of which 101,079 are direct (\ie 18.6\%).
We observe that \rubygems packages and external projects have more than thrice as many direct dependencies as \npm packages and external projects. This is in line with the observations made by Decan~\etal\cite{decan2019package}. \fig{fig:dependency_evolution} shows the evolution over time of the cumulative number of direct and indirect dependencies for \changed{package latest releases} and external projects considered in this study, grouped by package distribution.

\section{Empirical Analysis}
\label{sec:results}

Using the datasets of \sect{sec:method}, this section answers the research questions introduced in \sect{sec:intro}. %
For $RQ_0$ to $RQ_2$, we present statistical analyses based on the affected \npm and \rubygems package releases according to the vulnerability dataset.
For $RQ_3$ to \minor{$RQ_6$} we study the impact \changed{of affected releases on exposed packages and exposed projects that directly or indirectly depend on them.}

\changed{As part of the statistical analyses, %
we use the non-parametric Mann-Whitney U test to compare various types of distributions without assuming them to follow a normal distribution. The null hypothesis $H_0$ states that there is no difference between two distributions.} We set a global confidence level of 95\%, corresponding to a significance level of $\alpha = 0.05$. To achieve this overall confidence, the $p$-value of each individual test is compared against a lower $\alpha$ value, following a Bonferroni correction\footnote{If $n$ different tests are carried out over the same dataset, for each individual test one can only reject $H_0$ if $p< \frac{0.05}{n}$. In our case \changed{$n=48$, \ie $p<0.001$}.}.
\changed{If the null hypothesis can be rejected, we report the effect size with Cliff's delta $d$, a non-parametric measure that quantifies the difference between two populations beyond the interpretation of $p$-values.} Following the guidelines of Romano~\etal~\cite{romano2006exploring}, we interpret the effect size to be \emph{negligible} if $|d|\in[0,0.147[$, \emph{small} if $|d|\in[0.147,0.33[$, \emph{moderate} if $|d|\in [0.33,0.474[$ and \emph{large} if $|d|\in [0.474,1]$.

We use the technique of survival analysis~\cite{Klein2013} to estimate the probability that an event of interest will happen. Survival analysis creates a model estimating the survival rate of a population over time until the occurrence of an event, considering the fact that some subjects may leave the study, while for others the event of interest might not occur during the observation period. 
We rely on the non-parametric Kaplan-Meier statistic estimator commonly used to estimate survival functions.

All code used to carry out this analysis is available in a replication package~\footnote{\url{https://github.com/AhmedZerouali/vulnerability_analysis}}.\subsection{\rqZero}
\label{subsec:rq0}

This research question aims to characterise the \changed{vulnerability dataset of \sect{subsec:vulnerability}, its evolution over time as well as the number of package releases affected by these vulnerabilities.
After the identification of package releases and dependencies in \sect{subsec:dependency}, we found that 88 vulnerabilities do not affect any package release that is used as a dependency. Therefore, our final dataset contains 2,786 vulnerabilities, of which 2,118 directly affecting \npm packages and 668 directly affecting \rubygems packages.}

\begin{figure}[!ht]
	\begin{center}
		\setlength{\unitlength}{1pt}
		\footnotesize
		\includegraphics[width=0.98\columnwidth]{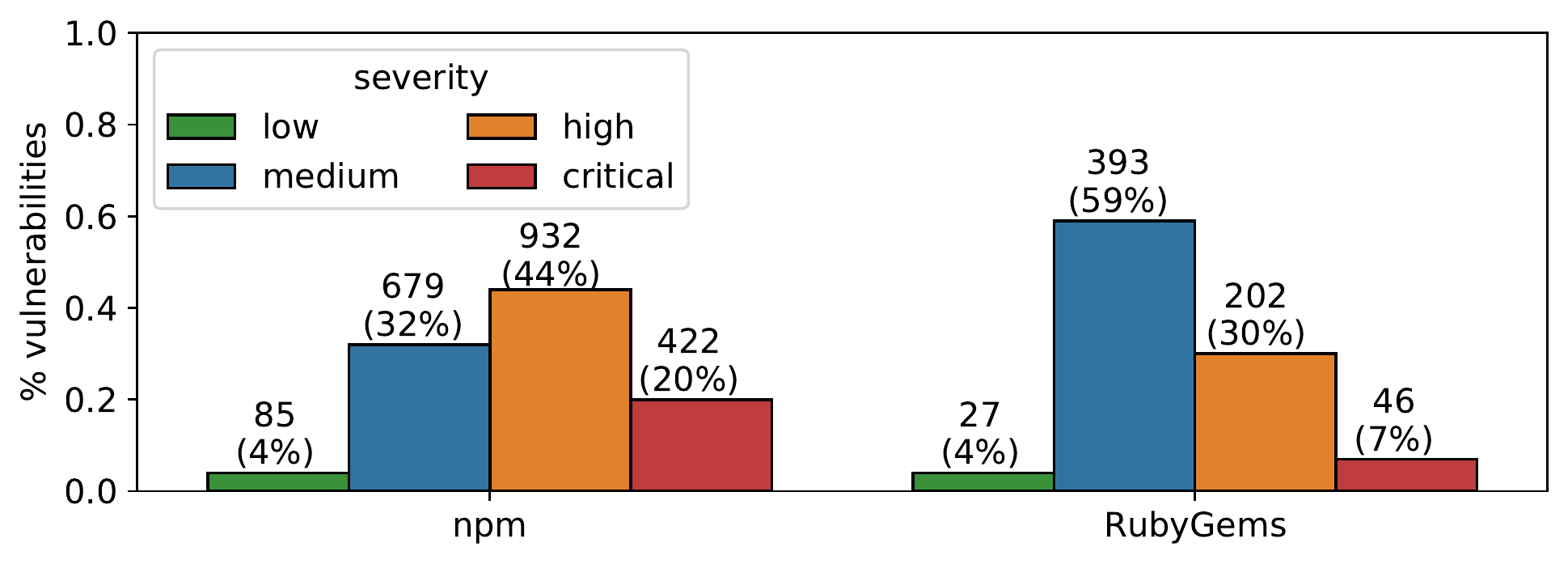}
		\caption{Proportion and number of considered vulnerabilities affecting \npm and \rubygems packages, grouped by severity.}
		\label{fig:number_vulns}
	\end{center}
\end{figure}

\fig{fig:number_vulns} depicts the number of considered vulnerabilities in this study, and the proportion of \textit{low, medium, high} or \textit{critical} severities for each package distribution. We observe that most of the vulnerabilities are of \textit{medium} or \textit{high} severity (76\% for \npm and 89\% for \rubygems). We also observe that \npm has nearly thrice as many \textit{critical} vulnerabilities as \rubygems. Another difference is that \npm has more \textit{high} vulnerabilities than \textit{medium} ones, while the inverse is true for \rubygems. The collected vulnerabilities affect 1,672 \npm and 321 \rubygems packages. The oldest vulnerability in \npm was \changed{disclosed} in June 2011, whereas the oldest one in \rubygems was \changed{disclosed} in August 2006. We also found that 1,175 (42\%) of the vulnerabilities did not have any known fix, of which 1,058 from \npm and 117 from \rubygems.

\minor{Finding the reasons behind the observed differences between both ecosystems in terms of number, severity, and types of vulnerabilities (see \tab{tab:vuln_names}) can be hard.} 
Each package distribution has its own tools, practices and policies \cite{Bogart2021} as well as differences in topological structure of the dependency network and size and growth of the ecosystem \cite{Decan2019}.
Both package distributions also focus on different programming languages.
\fig{fig:number_vulns} therefore does not normalize against the size of each ecosystem. The proportion of vulnerable packages is 0.22\% for \npm while it is 0.4\% for \rubygems, even though \npm is 8 times larger than \rubygems.

\begin{figure}[!ht]
	\begin{center}
		\setlength{\unitlength}{1pt}
		\footnotesize
		\includegraphics[width=0.98\columnwidth]{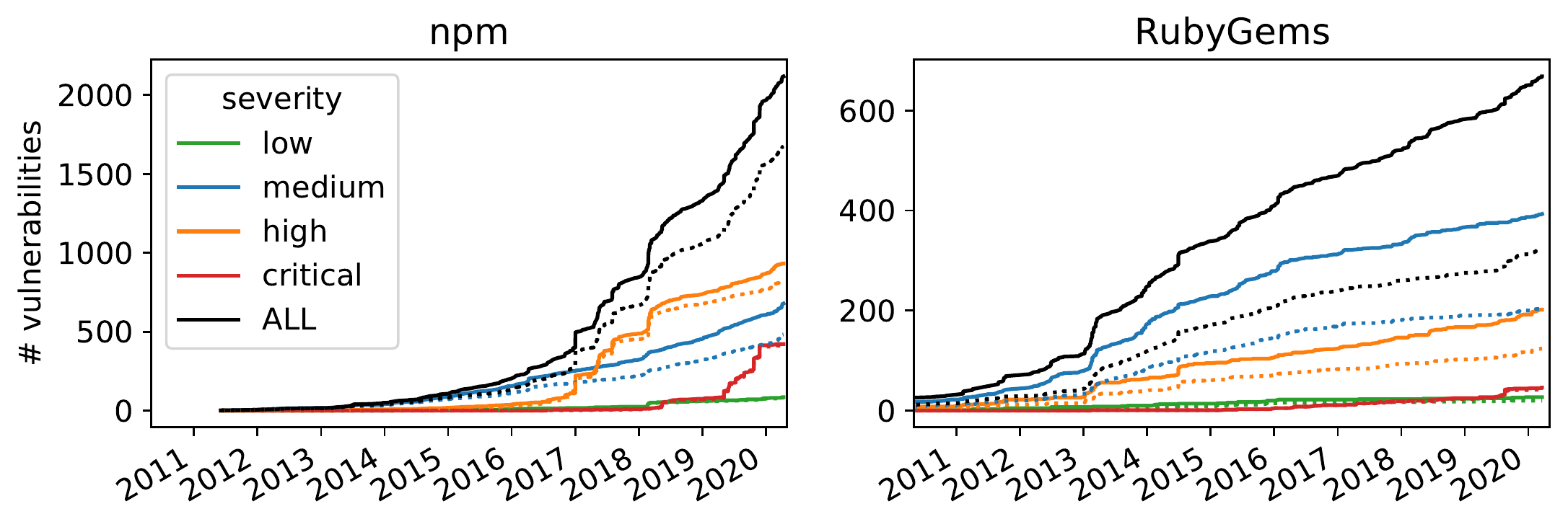}
		\caption{Temporal evolution of the cumulative number of \changed{disclosed} vulnerabilities (straight lines) and the corresponding number of affected packages (dotted lines) per severity.}
		\label{fig:affected_packages}
	\end{center}
\end{figure}

\fig{fig:affected_packages} depicts the evolution over time of the cumulative number of vulnerabilities (straight lines) grouped by severity, and their corresponding affected packages (dotted lines). The y-axis scale for \npm is different from the one for \rubygems because more vulnerabilities have been reported for \npm.
The number of reported vulnerabilities and thereby affected packages increases for both \npm and \rubygems over time. We also see that before 2017, the number of medium vulnerabilities was higher than the number of high vulnerabilities in \npm. Since 2016 and until 2018, the number of high vulnerabilities started increasing following a different trend. %

A similar trend can be observed for the critical vulnerabilities in 2019. This means that proportionally speaking, the severity of \npm vulnerabilities tends to increase over time, while such an observation cannot be made for \rubygems.

Looking at \textit{ALL} vulnerabilities we can see an exponential growth in the number of \npm vulnerabilities, while \rubygems shows a linear growth rate.
This is statistically confirmed by a regression analysis using linear and exponential growth models. The $R^2$ values reflecting the goodness of fit are summarized in \tab{tab:r_square_vuls}.~\footnote{$R^2 \in [0,1]$ and the closer to 1 the better the model fits the data.}
For both vulnerabilities and affected packages, \npm follows an exponential growth while \rubygems follows a linear one.
The exponential growth of \npm is in line with the exponential growth of its total number of packages\changed{~\cite{Decan2019}}. 
Given that \npm is by far the largest package distribution available\footnote{According to \librariesio, in May 2021, \npm contained 1.79M packages compared to ``only'' 173K packages in \rubygems.}, it is considerably more likely to \changed{contain reported vulnerabilities. \npm's popularity may attract more malicious developers on the one hand, that want to exploit vulnerabilities contained in some of its packages, and on the other hand it may attract more security researchers that aim to find and report vulnerabilities before they can be exploited.}

\begin{table}[!ht]
	\centering
	\caption{$R^2$-values of regression analysis on the evolution of the number of vulnerabilities and affected packages.}
	\label{tab:r_square_vuls}
	\begin{tabular}{l|r|r|r|r}

		\multirow{2}{*}{} & \multicolumn{2}{c|}{\bf \npm}   & \multicolumn{2}{c}{\bf \rubygems} \\
		& \bf \# vulns & \bf \# packages & \bf \# vulns   & \bf \# packages  \\ \hline
		\bf linear            & 0.85            & 0.84     & \underline{0.94}             & \underline{ 0.93}      \\
		\bf exponential       & \underline{0.96}            & \underline{0.97}     & 0.90               & 0.90
	\end{tabular}
\end{table}

For each vulnerability report in the vulnerability dataset, we identified the affected range of releases of the specified package in our package dataset.
Here too we relied on the constraint parser proposed in~\cite{decan2019package}.
While the vulnerability dataset corresponds to a relatively low number of vulnerable packages (see above), the majority of their releases is in fact affected: $67.4\%$ (\ie $43,330$ out of $64,236$) for \npm and $63.8\%$ (\ie $11,488$ out of $17,987$) for \rubygems.
$89.9\%$ of these affected releases (comprising both \npm and \rubygems) concern vulnerabilities of either \textit{medium} severity ($52.9\%$) or \textit{high} severity ($37\%$ \textit{high}).
Regardless of their severity, $57.3\%$ and $27.8\%$ of all \npm and \rubygems vulnerabilities are affecting more than 90\% of the releases of their corresponding packages. \changed{Most of these vulnerabilities (85.6\% and 62.9\%, respectively) are open ones that do not have any fix. Focusing only on fixed vulnerabilities, this proportion decreases to 16.2\% and 12.5\% for \npm and \rubygems vulnerabilities affecting more than 90\% of their package releases, respectively. This means, for the clients of these packages, that they should be using the latest available releases, especially the clients of \npm packages.}

\tab{tab:vuln_names} shows the top ten vulnerability types affecting \npm and \rubygems packages, with the number of vulnerabilities of each type grouped by severity. 
We observe that the most prevalent vulnerability type is \emph{Malicious Package} and \emph{Cross-site Scripting (XSS)} found in 19.3\% of all \npm vulnerability reports, and 17.4\% of all \rubygems vulnerability reports. We observe that \npm and \rubygems packages are exposed to similar vulnerabilities but with different occurrences. Some vulnerability types seem to affect \javascript packages more than \ruby packages, \eg \emph{Malicious Package} and \emph{Directory Traversal}.

\begin{table}[!ht]
	\centering
\caption{Top ten vulnerability types affecting \npm and \rubygems packages, with the number of vulnerabilities of each type grouped by severity (C~=~\critical, H~=~\high, M~=~\medium, L~=~\low).}
\label{tab:vuln_names}
\begin{tabular}{lr|rrrr}
	\bf \npm vulnerability types & \bf \#vulns &   \bf C & \bf H & \bf M & \bf L \\
	\hline
	Malicious Package  &  410                           & 345 & 64 & 1 & 0 \\
	Directory Traversal &  331                          & 5 & 297 & 28 & 1\\
	Cross-site Scripting  & 322                    		& 15 & 55 & 245 & 7 \\
	Resource Downloaded over Insecure Protocol &   154	 & 0 & 145 & 8 & 1\\
	Regular Expression Denial of Service  &  138 		& 0 & 57 & 50 & 31\\
	Denial of Service   &   91                    		 & 5 & 52 & 32 & 2\\
	Prototype Pollution  &   77                         & 1 & 38 & 36 & 2\\
	Command Injection &   57                            & 0 & 20 & 34 & 3\\
	Arbitrary Code Execution  &   44                    & 11 & 24 & 7 & 2 \\
	Arbitrary Code Injection &   36                     & 7 & 9 & 20 & 0 \\

	\\ \bf \rubygems vulnerability types & \bf \#vulns &   \bf C & \bf H & \bf M & \bf L \\
	\hline
	Cross-site Scripting  &  116 			& 0 & 6 & 108 & 2 \\
	Denial of Service  &   54 				& 1 & 18 & 34 & 1 \\
	Arbitrary Command Execution &   47  	& 2 & 29 & 16 & 0\\
	Information Exposure &   40 			& 1 & 6 & 27 & 6 \\
	Arbitrary Code Execution &   27 		& 2 & 16 & 9 & 0\\
	Man-in-the-Middle  &   26 				& 0 & 3 & 22 & 1\\
	Malicious Package &   19 				& 18 & 1 & 0 & 0\\
	Cross-site Request Forgery &   18 		& 0 & 4 & 14 &  0\\
	SQL Injection &   18 					& 1 & 11 & 6 & 0\\
	Directory Traversal &   16 				& 0 & 5 & 10 & 1\\
\end{tabular}
\end{table}

\smallskip
\noindent\fbox{%
	\parbox{0.98\textwidth}{%
	The number of reported vulnerabilities is increasing exponentially for \npm and linearly for \rubygems.
	The relative proportion of high and critical vulnerabilities seems to increase over time.
	Two out of three releases of vulnerable packages are affected by at least one vulnerability for both ecosystems.
	The types of most prevalent vulnerabilities differ between \npm and \rubygems.
	}%
}
\subsection{\rqOne}
\label{subsec:rq1}
Delayed fixing of security vulnerabilities puts software packages and their dependents at risk as it lengthens the window that hackers have to discover and exploit the vulnerability. Unknown vulnerabilities may linger and remain to be present in more recent releases of a vulnerable package, exposing the dependents of an entire release range for a substantial amount of time.
For example, in January 2021, a vulnerability named ``Baron Samedit" was discovered in the popular Linux package {\sf sudo}~\footnote{\url{https://snyk.io/vuln/SNYK-DEBIAN9-SUDO-1065095}}. This vulnerability allowed local users to gain root-level access and was introduced in the {\sf sudo} code back in July 2011, effectively exposing all releases during the past decade and therefore millions of deployments. 

\begin{figure}[!ht]
	\begin{center}
		\setlength{\unitlength}{1pt}
		\footnotesize
		\includegraphics[width=0.98\columnwidth]{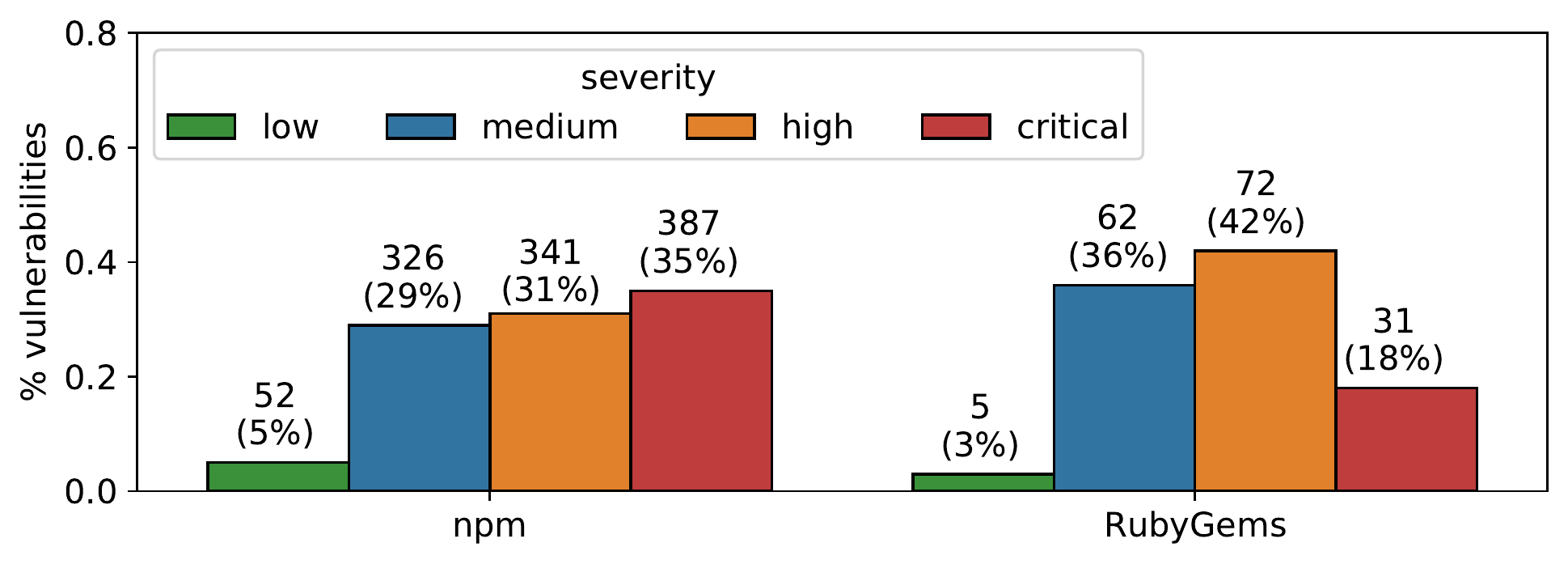}
		\caption{Proportion and number of vulnerabilities after filtering out old vulnerabilities and inactive packages.}
		\label{fig:number_vulns_active}
	\end{center}
\end{figure}

\changed{$RQ_1$  aims to study how much time it takes until a vulnerability is discovered. Since it is not possible to accurately know when a vulnerability was actually discovered, we rely instead on the vulnerability \emph{disclosure} date as a proxy for the \emph{discovery} date and we study the time needed before a lingering vulnerability gets disclosed.
	
Usually, when a vulnerability is discovered, a CVE identifier will be reserved for it so it can be uniquely identified later. These CVE IDs can be reserved by a Central Numbering Authorities (CNAs) that have the priority to reserve CVE IDs from MITRE~\footnote{\url{https://cve.mitre.org/cve/request_id.html}}.
\snyk, \github and HackerOne are examples of such CNAs. To study the disclosure time in more detail, we inspected MITRE using the CVE identifiers we obtained for the vulnerabilities in our vulnerability dataset. We extracted the reservation date and CNA of each CVE and found that only 1,487 vulnerabilities (55\%) have a CVE linked to them. 820 of them were disclosed before the CVE reservation date, 639 were disclosed after that date, and 100 vulnerabilities had a disclosure date that coincided with the CVE reservation date. We found many organizations responsible for reserving the CVEs of vulnerabilities in \npm and \rubygems packages, including Red Hat, Microsoft and GitHub. The top three of CNAs with the most CVEs was HackerOne, Mitre and Snyk for \npm, and Mitre, Red Hat and HackerOne for \rubygems.

For all vulnerabilities, we computed the \emph{disclosure lag}, \ie the number of days between the first affected release and the vulnerability \changed{disclosure} date. \fig{fig:discovery_evolution} shows the evolution over time of the disclosure lag \minor{distribution}, grouped by CNA. \minor{A gap can be observed between 2018 and 2020 for \textit{RubyGems - NO CVE} because during this period we did not find any disclosed vulnerability for \rubygems with the NO CVE id.}
The disclosure lag tends to increase in both package distributions, implying that recently reported vulnerabilities take longer to disclose than older ones. This trend can be observed for all CNAs. We also observe a longer disclosure lag for \rubygems vulnerabilities than for \npm ones.
Comparing between CNAs, \fig{fig:discovery_evolution} also reveals that some CNAs tend to disclose and report vulnerabilities faster than others.

\begin{figure}[!ht]
	\begin{center}
		\setlength{\unitlength}{1pt}
		\footnotesize
		\includegraphics[width=0.98\columnwidth]{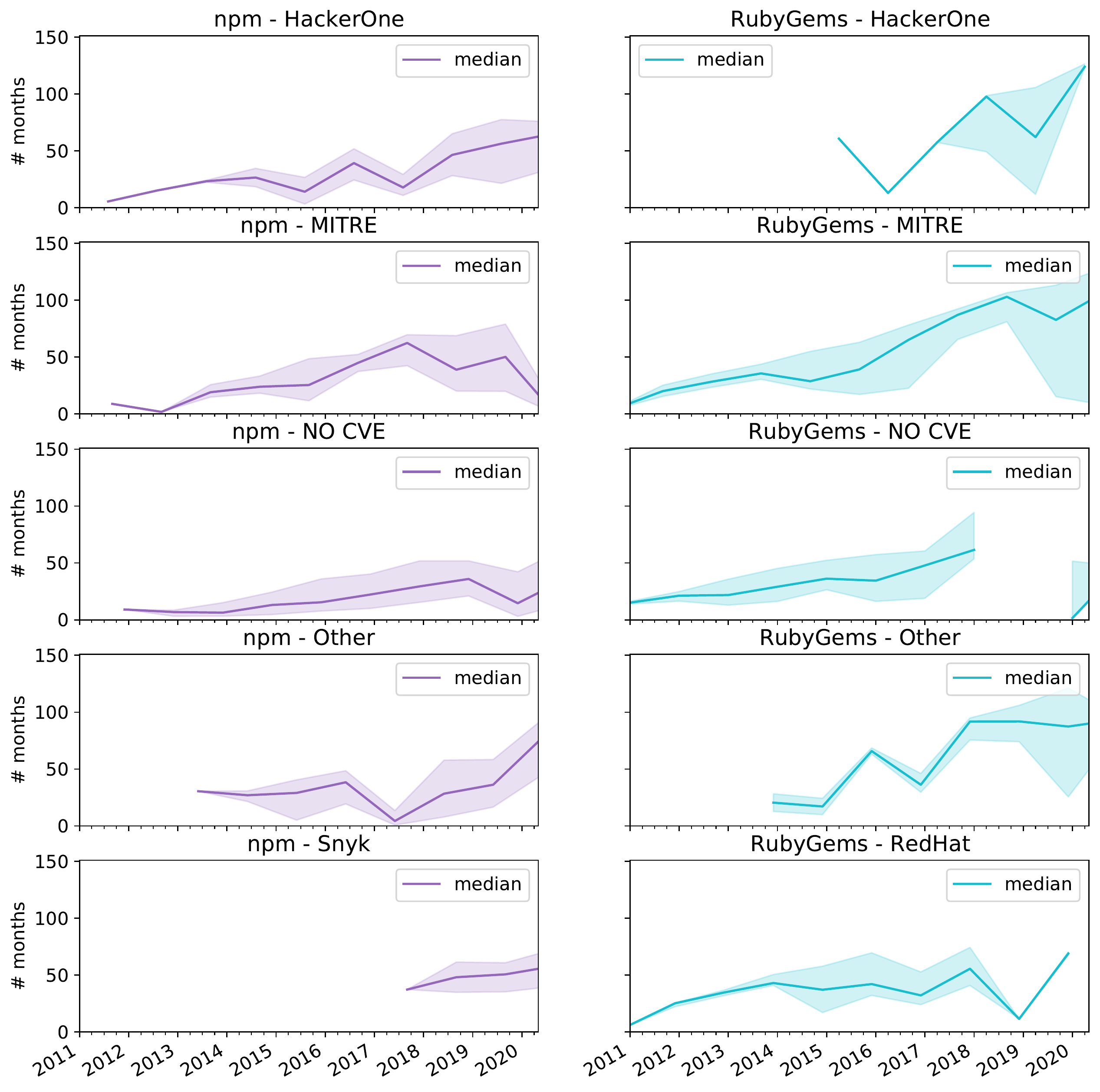}
		\caption{Evolution of the vulnerability disclosure lag \minor{distribution} in \npm and \rubygems, grouped by CVE Central Numbering Authority. The shaded areas represent the interval between the $25^{th}$ and $75^{th}$ percentile.}
		\label{fig:discovery_evolution}
	\end{center}
\end{figure}

To make a fair comparison, we decided to focus on recent vulnerabilities that have been disclosed in the last three years, \ie after 2017-04-17. 
In addition, we focus only on vulnerabilities that have been disclosed in packages that received at least one update in the last two years.
Indeed, since inactive packages have less maintenance activity, it would be unfair to compare vulnerabilities disclosed in these packages with vulnerabilities of packages receiving continuous maintenance, including vulnerability inspection.
This led us to consider only 1,276 vulnerabilities (45.8\%) for $RQ_1$ and $RQ_2$. \fig{fig:number_vulns_active} shows the proportion and number of vulnerabilities kept after this filtering. Compared to the unfiltered set in \fig{fig:affected_packages} there is a considerably higher proportion of critical vulnerabilities.
}

\begin{figure}[!ht]
	\begin{center}
		\setlength{\unitlength}{1pt}
		\footnotesize
		\includegraphics[width=0.98\columnwidth]{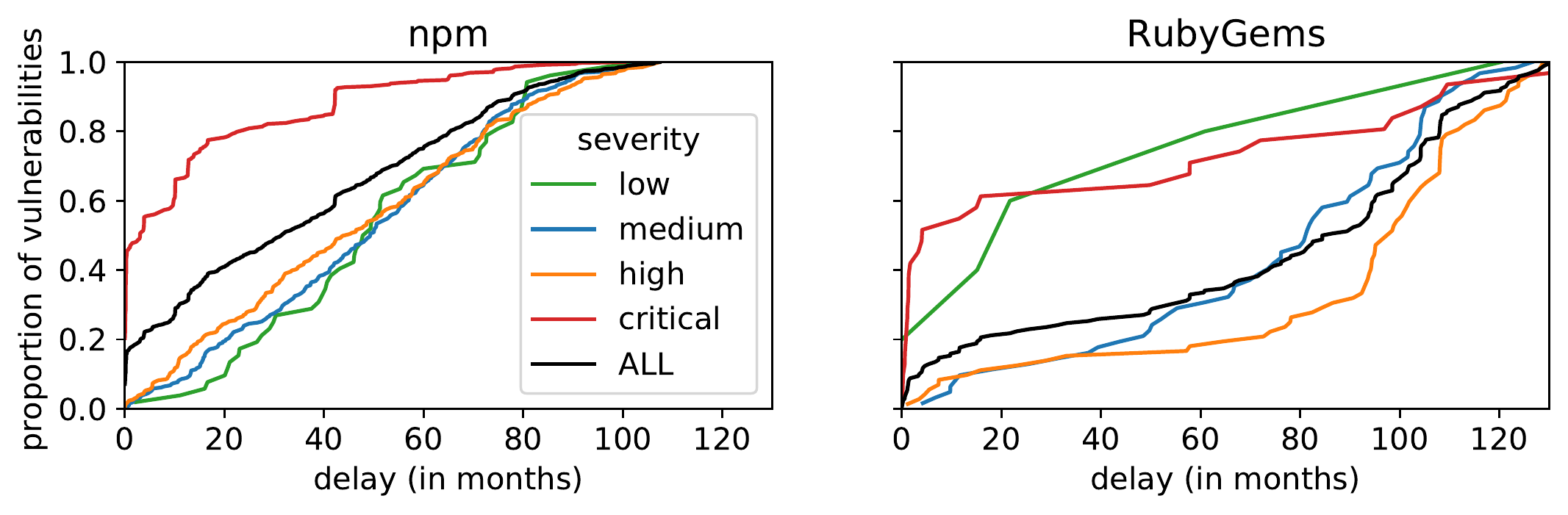}
		\caption{Cumulative proportion of \changed{disclosed} vulnerabilities in function of the time elapsed since the first affected package release, grouped by severity level.}
		\label{fig:discovery_proportion}
	\end{center}
\end{figure}

\fig{fig:discovery_proportion} shows the cumulative proportion of \changed{disclosed vulnerabilities and their disclosure lag, grouped by severity level. For \npm, we observe that critical vulnerabilities are the fastest to disclose. It only takes 3.1 months to disclose 50\% of all \textit{critical} vulnerabilities, while it takes respectively 49.3, 49.3 and 44.5 months to disclose 50\% of all \textit{low}, \textit{medium} and \textit{high} vulnerabilities.
For \rubygems, the disclosure lag seems to depend much less on its severity level. 
Using log-rank tests for both package distributions, we could only confirm a statistically significant difference for the comparisons with \textit{critical} vulnerabilities in \npm (\eg \textit{critical} vs \textit{high}). %
For all other comparisons in \rubygems, the null hypothesis stating that there is no difference in disclosure lag could not be rejected.

Disregarding the severity status, vulnerabilities have a lower disclosure lag for \npm than for \rubygems (in \npm it takes 31.5 months to disclose 50\% of all vulnerabilities compared to 84.4 months for \rubygems). To statistically confirm this observed difference in disclosure lag, we carried out a Mann-Whitney U test. The null hypothesis could be rejected with a \textit{large} effect size ($|d|=0.53$), statistically confirming that vulnerabilities take longer to disclose for \rubygems than for \npm. This could also be observed when grouping the analysis by CNA. Possible explanations for this finding may be that \npm has better security detection tools and more security researchers than \rubygems. We found that vulnerabilities of \npm packages were disclosed by 635 security researchers and communicated to MITRE via 21 CNAs, while \rubygems vulnerabilities were disclosed by 236 security researchers and communicated to MITRE via 17 CNAs.

Redoing the same analysis for \emph{inactive} libraries, we found that it took 24.8 and 34.3 months to disclose 50\% of the vulnerabilities in inactive \npm and \rubygems packages, respectively. To compare between inactive and active packages, we carried out Mann-Whitney U tests. We could only find a statistically significant difference with a \textit{large} effect size ($|d|=0.48$) in favor of active packages when doing the comparison for \rubygems. This means that, for \rubygems, vulnerabilities of inactive packages took less time to be disclosed than those of active packages.

Since \textit{Malicious Package} vulnerabilities are intentionally injected in the form of new updates or packages, we compared them to other vulnerabilities and expected them to be disclosed faster. Indeed, we found that 50\% of the \textit{Malicious Package} vulnerabilities in \npm are disclosed within 12 days, while those of \rubygems are disclosed within 35 days. Carrying out Mann-Whitney U tests between \textit{Malicious Package} vulnerabilities and other vulnerabilities, we could only find a statistically significant difference with a \textit{small} effect size ($|d|=0.25$) in favor of non-\textit{Malicious Package} vulnerabilities when doing the comparison for \npm. This means that, for \npm, \textit{Malicious Package} vulnerabilities are disclosed faster than other vulnerabilities.}

\smallskip
\noindent\fbox{%
	\parbox{0.98\textwidth}{%
	In \npm, \changed{critical vulnerabilities are disclosed faster.}
	Vulnerabilities in \npm are \changed{disclosed} faster than in \rubygems.
	It takes \changed{2.3 and 7 years to disclose half of the vulnerabilities} lingering in \npm and \rubygems packages, respectively. The \changed{vulnerability disclosure lag} has been increasing over time in both package distributions. \changed{In \npm, Malicious Package vulnerabilities are disclosed faster than other vulnerability types.}
	}%
}
\subsection{\rqTwo}
\label{subsec:rq2}

\begin{figure}[!ht]
	\begin{center}
		\setlength{\unitlength}{1pt}
		\footnotesize
		\includegraphics[width=0.98\columnwidth]{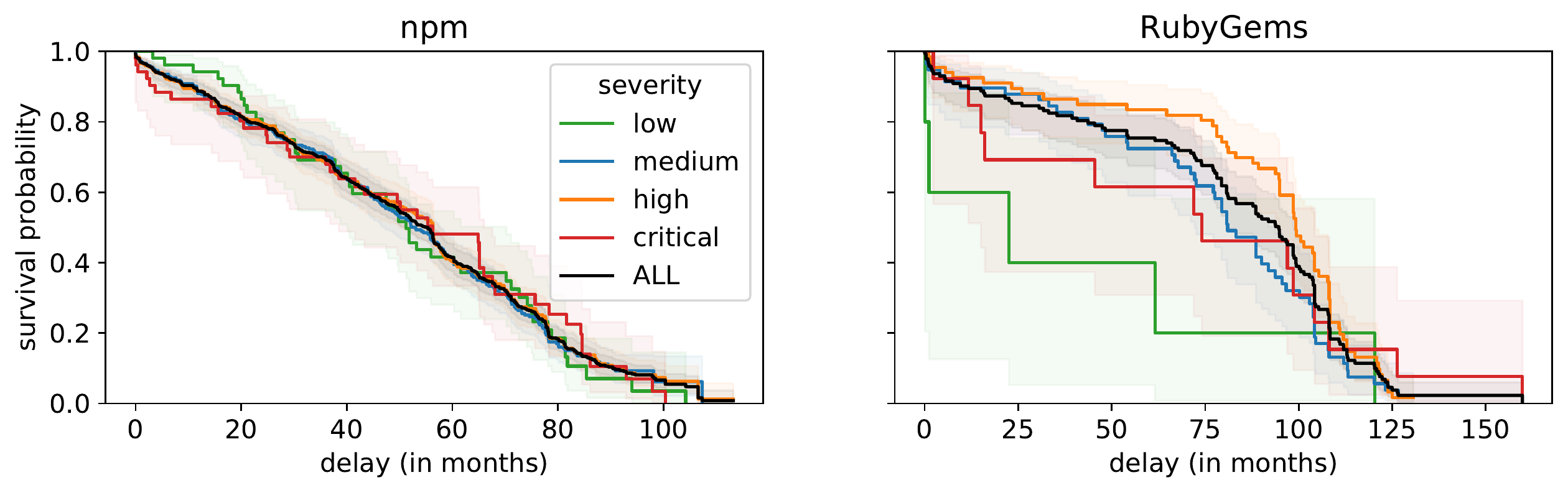}
		\caption{Survival probability for event ``vulnerability is fixed'' since the first affected release. The shaded colored areas represent the confidence intervals ($\alpha=0.05$) of the survival curves.}
		\label{fig:survival_rq2}
	\end{center}
\end{figure}

\changed{$RQ_1$ studied the \emph{disclosure lag} between the first package release affected by a vulnerability and the moment this vulnerability was disclosed. 
$RQ_2$ investigates \minor{the time that a vulnerability remains in a package until its fix (1) since the first affected release and (2) since the disclosure time. Case (1) refers to the number of days between the release date of the first affected release and the release date of the first package update that is no longer affected by the vulnerability, while case (2) refers to the number of days between the disclosure date of the vulnerability and the release date of the first package update that is no longer affected by the vulnerability.}
This \minor{analysis} is relevant since the longer a package remains affected, the longer it will remain a source of vulnerabilities to potential users and dependents. The latter will be obliged to rely on vulnerable package releases as long as the package maintainers did not release a package update that fixes the vulnerability. This also harms the package dependency network since if a vulnerability fix is being delayed, more and more dependents may potentially make use of the vulnerable package or may become potentially exposed through transitive dependencies. The later a fix becomes available, the more difficult it becomes for all dependents to update their transitive dependencies.

$RQ_2$ uses the same filtered dataset as $RQ_1$, ignoring inactive packages and focusing only on recent vulnerabilities.
In addition, we exclude \textit{Malicious Package} vulnerabilities since they are known to be fixed in a different way, most frequently by simply removing the malicious package (release) from the registry.

\minor{The analysis of $RQ_2$ will be subdivided into three parts: $RQ_2^a$ a characterisation of the type of versions in which vulnerabilities are fixed; $RQ_2^b$ the time between the first affected release and the fix; and $RQ_2^c$ the time between the vulnerability disclosure and the fix.}

\minor{\textbf{\subsubsection*{$RQ_2^a$ Version types in which vulnerabilities are fixed.}}}

Relying on the \semver specification~\cite{preston2013semantic,decan2019package}, we studied the \changed{package release version (\ie patch, minor or major) that fixes a vulnerability. The purpose is to determine whether there is a relation between the severity of the vulnerability and the version type of the first release that included a fix. According to \semver, patch releases are the most likely candidates for vulnerability fixes.}

As both \npm and \rubygems promote and encourage the use of \semver, we expect to find most of their vulnerable packages to be fixed in a patch release.
\fig{fig:releases_rq2} presents stacked bar plots showing the proportion of fixed vulnerabilities per severity, \changed{grouped by the version type of the first unaffected release.\footnote{We implicitly assume here that the first unaffected release is the one containing the fix.} Our expectations are confirmed, since the majority of vulnerabilities are fixed in patch releases. Disregarding the severity level, 65\% of the vulnerabilities were fixed in patch releases, 22.2\% in minor releases and only 12.8\% in major releases. The severity of a vulnerability does not seem to play a major role in the version type of the release that contains its fix.  An exception are the \textit{low} vulnerabilities in \rubygems: in contrast to \npm, they do not seem to be considered as important, as they are fixed more often in minor and major releases that incorporate other types of changes as well.}

\begin{figure}[!ht]
	\begin{center}
		\setlength{\unitlength}{1pt}
		\footnotesize
		\includegraphics[width=0.98\columnwidth]{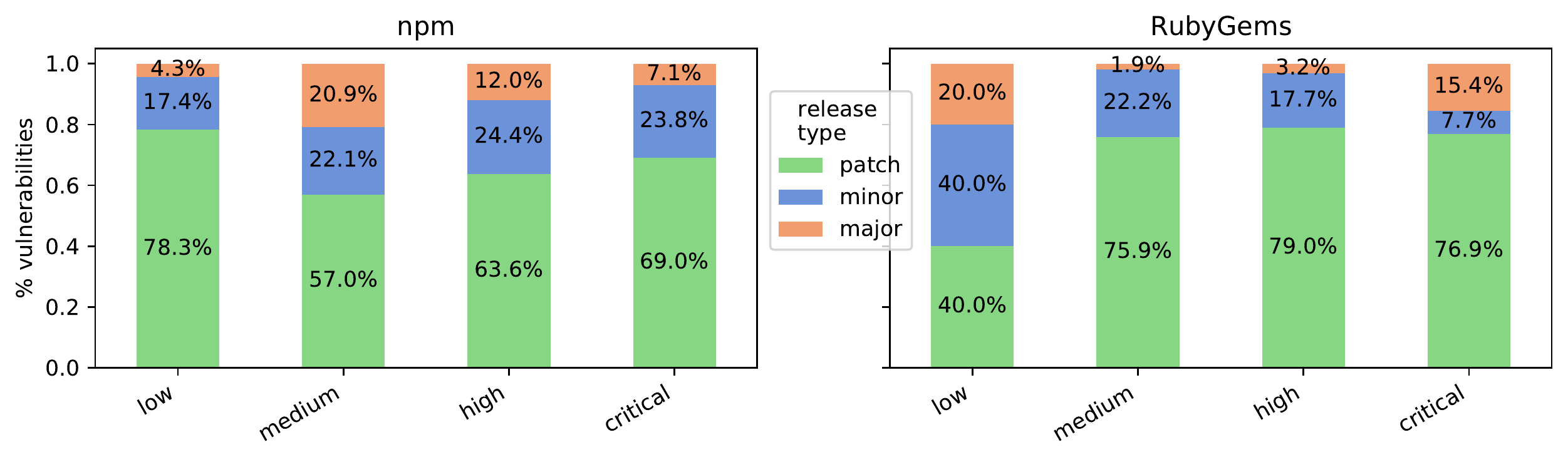}
		\caption{Proportion of fixed vulnerabilities per severity, grouped by version type of the first unaffected release.}
		\label{fig:releases_rq2}
	\end{center}
\end{figure}

\smallskip
\noindent\fbox{%
	\parbox{0.98\textwidth}{%
		\changed{65\% of all \minor{disclosed} vulnerabilities are fixed in patch releases. \textit{Low} severity vulnerabilities in \rubygems are fixed less often in patch releases. The severity of a vulnerability does not seem to have an impact on the first release type in which the vulnerability is fixed.}
	}%
}
\smallskip

\minor{\textbf{\subsubsection*{$RQ_2^b$ Time between the first affected release and the fix.}}}

Since many vulnerabilities in our dataset did not yet receive a fix (128 out of 693 for \npm, and 9 out of 143 for \rubygems), we used a survival analysis~\cite{Klein2013} to estimate the probability over time for the event ``vulnerability is fixed" with respect to the date of the first affected release, and grouped by severity type. \fig{fig:survival_rq2} shows the Kaplan-Meier survival curves for this analysis. For \npm, the confidence intervals of all survival curves overlap, suggesting that there is no difference in \minor{the time to fix a vulnerability since its first appearance} depending on the severity of the vulnerability.
A similar observation can be made for \rubygems, with the exception of \textit{low} severity vulnerabilities that seem to be fixed considerably faster than any of the other severity types.
It takes only 22 months to fix 50\% of all \textit{low} vulnerabilities in \rubygems, while this is 81, 99 and 74 months for, \textit{medium}, \textit{high} and \textit{critical} vulnerabilities. 
However, log-rank tests did not allow us to confirm any statistical differences between severity levels for any package distribution.

When ignoring the severity level, however, a log-rank test could confirm a statistically significant difference in fixing \minor{time (since the first affected release)} between \npm and \rubygems. Vulnerabilities in \npm have a considerably smaller fixing time. For example, half of all vulnerabilities in \npm are fixed after 55 months, while it takes 94 months in \rubygems.}

\changed{The fairly long observed time between \minor{the first affected release and the fix} suggests that vulnerabilities affect many releases before they are fixed. Indeed, focusing only on vulnerabilities that have received a fix, an \npm vulnerability affects a median of 30 package releases before it is fixed. While for \rubygems, it affects a median of 59 releases.
The boxen plots of \fig{fig:affected_versions} show how many releases of a vulnerable package are affected by each severity level. We observe that the number of affected releases per vulnerable package is quite high. We also observe that for \rubygems, \textit{low} severity vulnerabilities affect less releases than other vulnerabilities. This is in line with the observations made in \fig{fig:survival_rq2}.

\begin{figure}[!ht]
	\begin{center}
		\setlength{\unitlength}{1pt}
		\footnotesize
		\includegraphics[width=0.98\columnwidth]{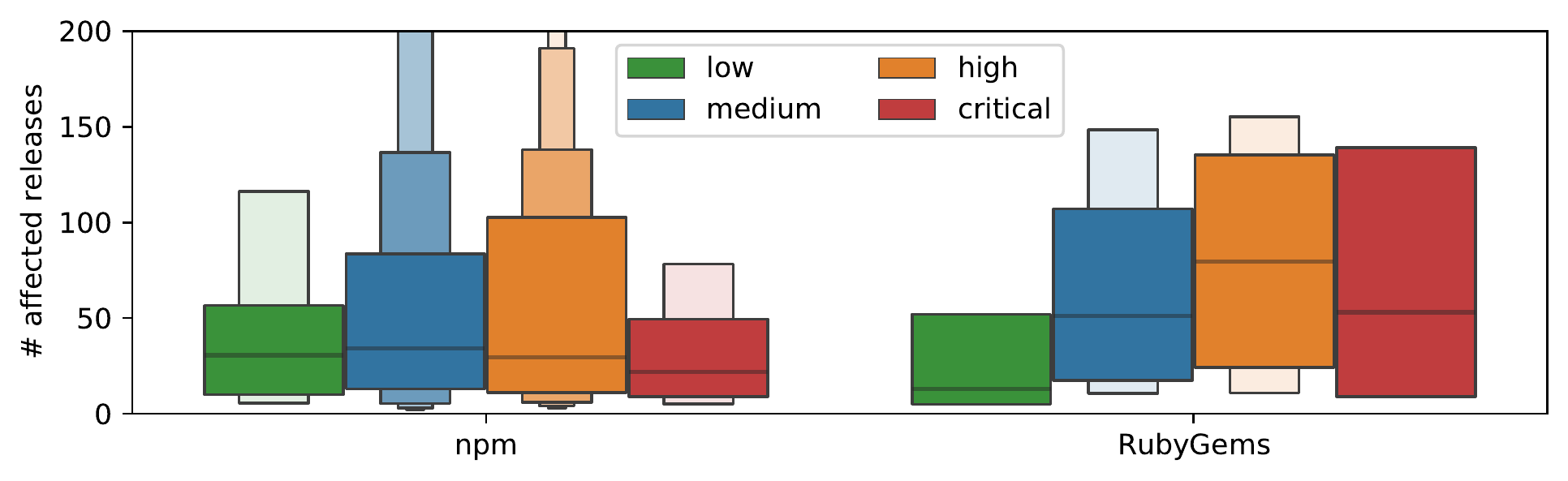}
		\caption{Boxen plots of the distribution of the number of affected releases of vulnerable packages in \npm and \rubygems, grouped by severity.}
		\label{fig:affected_versions}
	\end{center}
\end{figure}

\smallskip
\noindent\fbox{%
	\parbox{0.98\textwidth}{%
	\minor{Disclosed} vulnerabilities in \npm~\minor{take a considerably shorter time to fix since the first affected release}. Half of all \minor{disclosed} \npm vulnerabilities take 55 months to fix since their introduction, compared to 94 months for  \minor{disclosed} \rubygems vulnerabilities. As a consequence, the impact of vulnerabilities is higher for \rubygems, affecting a median of 59 package releases compared to 30 package releases for \npm.
}%
}
}

\smallskip

\minor{\textbf{\subsubsection*{$RQ_2^c$ Time between the vulnerability disclosure and the fix.}}}

\minor{It is important to fix vulnerabilities rapidly after their discovery, and especially after they have been publicly disclosed. If a vulnerability is publicly disclosed before a fix is available, more attackers will know about it and will be able to exploit it. Hence, when an open source vulnerability is reported to a security monitoring service, it is usually first disclosed privately in order to give the maintainers time to fix it before it is made public. For example, when \snyk receives a report about a vulnerable package, it informs the package maintainers and gives them 90 days to issue a remediation of the vulnerability. An extension can be granted at the maintainers' request, depending on the severity of the discovered vulnerability. To investigate whether maintainers fix their vulnerabilities within 90 days, we computed the time difference between the vulnerability fix date and the date when it was disclosed~\footnote{This analysis included Malicious Package vulnerabilities}. We found a significant proportion of vulnerabilities that exceeded this 90 day period: 17.8\% for \npm and 10\% for \rubygems. 

According to a survey carried out by \snyk in 2017, \textit{``$34\%$ of maintainers said they could respond to a security issue within a day of learning about it, and $60\%$ said they could respond within a week"~\cite{snyk2017}}. Our own quantitative observations align with this claim:
For \npm, 36.9\% of all vulnerabilities with a known fix were fixed within a day and 54.4\% were fixed within a week, while for \rubygems 46.2\% were fixed within a day and 63\% were fixed within a week. Overall, 38.9\% of all vulnerabilities with a known fix were fixed within a day, and 56.3\% were fixed within a week after their disclosure.

\smallskip
\noindent\fbox{%
	\parbox{0.98\textwidth}{%
		For \npm, 17.8\% of the fixed vulnerabilities needed more than 90 days after their disclosure to be fixed, while this proportion is 10\% for \rubygems. 38.9\% of all fixed vulnerabilities were fixed within a day, and 56.3\% were fixed within a week after their disclosure. 
	}%
}

}\subsection{\rqThree}
\label{subsec:rq3}
We have so far only studied vulnerable package releases.
\changed{$RQ_3$ focuses on packages as well as external projects that may} be \emph{exposed} to a vulnerability within their direct or indirect dependencies.
This exposure may lead to a security breach if the affected functionality of the vulnerable dependency is being used.
$RQ_3^a$ will study the exposure of \emph{\minor{the latest} package releases} to direct or indirect vulnerable dependencies, whereas
$RQ_3^b$ will study the exposure of \emph{external projects} to such vulnerable dependencies.
\subsubsection*{\textbf{\rqThreeA}}

For all 842,697 of the latest package releases available in the \npm and \rubygems snapshots, we determined the direct and indirect dependencies by resolving their dependency constraints.
We narrowed down the analysis to those dependencies that are referenced in the vulnerability dataset.
42.1\% of all considered \npm{} \minor{packages} (315,315 out of 748,026) and 39\% of all considered \rubygems{} \minor{packages} (36,957 out of 94,671) were found to have at least one vulnerable direct or indirect dependency \minor{in their latest release}. More specifically, 15.7\% of \npm{} \minor{latest} releases and 17.8\% of \rubygems{} \minor{latest} releases are directly exposed, while 36.5\% of \npm{} releases and 27.1\% of \rubygems releases are indirectly exposed~\footnote{The two categories of directly and indirectly exposed package releases are non-exclusive.}. \changed{This is in line with the findings of Zimmerman ~\etal~\cite{zimmermann2019small} reporting that up to 40\% of all packages depend on code with at least one publicly known vulnerability.} 

\smallskip
\noindent\fbox{%
	\parbox{0.98\textwidth}{%
	More than 15\% of the \minor{(latest)} dependent package releases are exposed to vulnerable {\bf direct} dependencies.
	\changed{36.5\% of \npm and 27.1\% of \rubygems latest package releases are exposed to vulnerabilities coming from vulnerable {\bf indirect} dependencies.}
	}%
}

\paragraph{\textbf{Vulnerable direct dependencies.}}

\begin{figure}[!ht]
	\begin{center}
		\setlength{\unitlength}{1pt}
		\footnotesize
		\includegraphics[width=0.98\columnwidth]{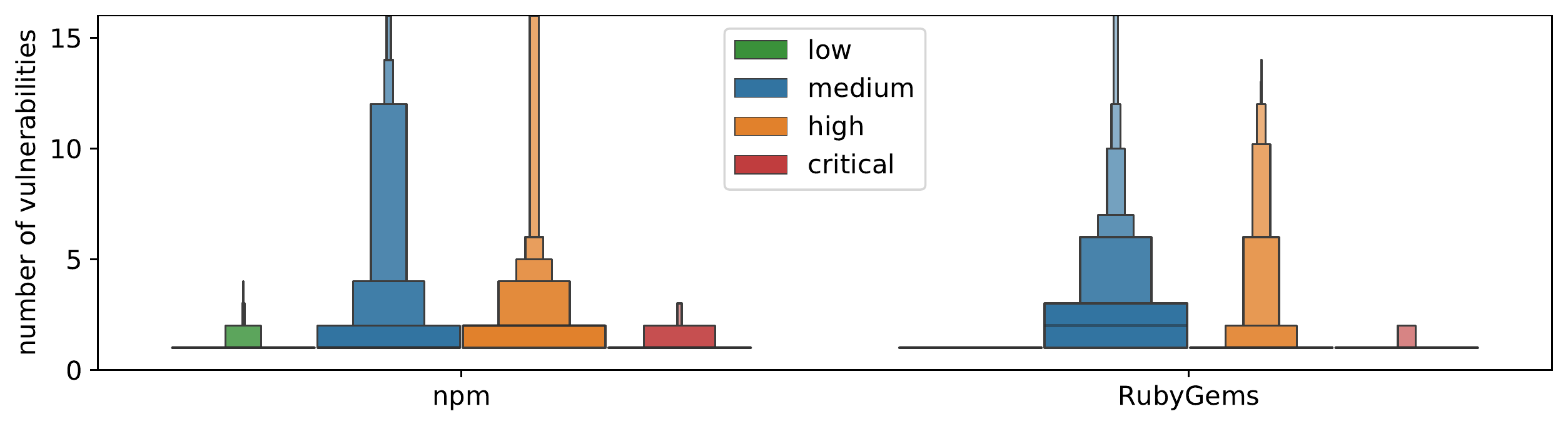}
		\caption{Boxen plots showing the distribution of the number of vulnerabilities in vulnerable {\bf direct} dependencies of the \npm and \rubygems snapshots, grouped by severity.}
		\label{fig:direct_deps}
	\end{center}
\end{figure}

We found only a small minority of direct dependencies (\ie package releases on which at least one other package directly depends) to be vulnerable (because they contain at least one vulnerability).
Of all 3,638,361 dependencies considered for \npm, only 154,455 (4.2\%) are vulnerable; and of all 224,959 dependencies considered for \rubygems, only 20,336 (9\%) are vulnerable.
\fig{fig:direct_deps} shows the distribution, per severity category, of the number of vulnerable package releases found among the direct dependencies in \npm and \rubygems.
We observe that medium and high severity vulnerabilities are the most common among vulnerable dependencies.
We also observe that vulnerable dependencies in \rubygems tend to be more often of medium severity than in \npm (blue boxen-plot), and less often of high severity (orange boxen-plot).
To statistically confirm these observed differences between \npm and \rubygems, we carried out Mann-Whitney U tests to compare the number of vulnerabilities per severity type. The null hypothesis could be rejected for all comparisons. %
However, the effect size (shown in \tab{tab:rq3_severities}) was {\em negligible} for all comparisons, except for the category of high severity vulnerabilities where a {\em moderate} effect size was reported.
Disregarding the severity category, the effect size was {\em small} in favour of \npm, suggesting that direct dependencies in \npm tend to have more vulnerabilities.

\begin{table}[!ht]
	\centering
\caption{Mean and median number of vulnerabilities found in direct dependencies, in addition to effect sizes and their directions when comparing \npm and \rubygems dependency vulnerabilities.}
\label{tab:rq3_severities}
	\begin{tabular}{l|rr|rr|ccr}
		\multirow{2}{*}{} & \multicolumn{2}{c|}{\bf\npm} & \multicolumn{2}{c|}{\bf\rubygems} & \multirow{2}{*}{\bf direction} & \multirow{2}{*}{ $|\textbf{d}|$} & \multirow{2}{*}{\bf effect size}\\
		& \bf mean       & \bf median      & \bf mean         & \bf median         &                            &                              \\ 
		\toprule
		\bf \low              & 1.07       & 1           & 1.02         & 1              & \textgreater{}                & 0.05    & negligible                     \\
		\bf \medium            & 2.55       & 1           & 2.73         & 2              & \textless{}                & 0.06                & negligible          \\
		\bf \high              & 2.32       & 2           & 1.81         & 1              & \textgreater{}             & 0.39            & {\bf moderate}              \\
		\bf \critical          & 1.17       & 1           & 1.04         & 1              & \textgreater{}                & 0.11           & negligible               \\ \hline
		\bf all               & 2.22       & 1           & 1.98         & 1              & \textgreater{}             & 0.16 & small 
	\end{tabular}
\end{table}

\smallskip
\noindent\fbox{%
	\parbox{0.98\textwidth}{%
	\rubygems has more than twice the  proportion of vulnerable direct dependencies than \npm  (9\% compared to 4.2\%).
	On the other hand, direct dependencies in \npm tend to have more vulnerabilities than direct dependencies in \rubygems.
	}%
}

\bigskip
\paragraph{\textbf{Vulnerable indirect dependencies.}}

\begin{figure}[!ht]
	\begin{center}
		\setlength{\unitlength}{1pt}
		\footnotesize
		\includegraphics[width=0.98\columnwidth]{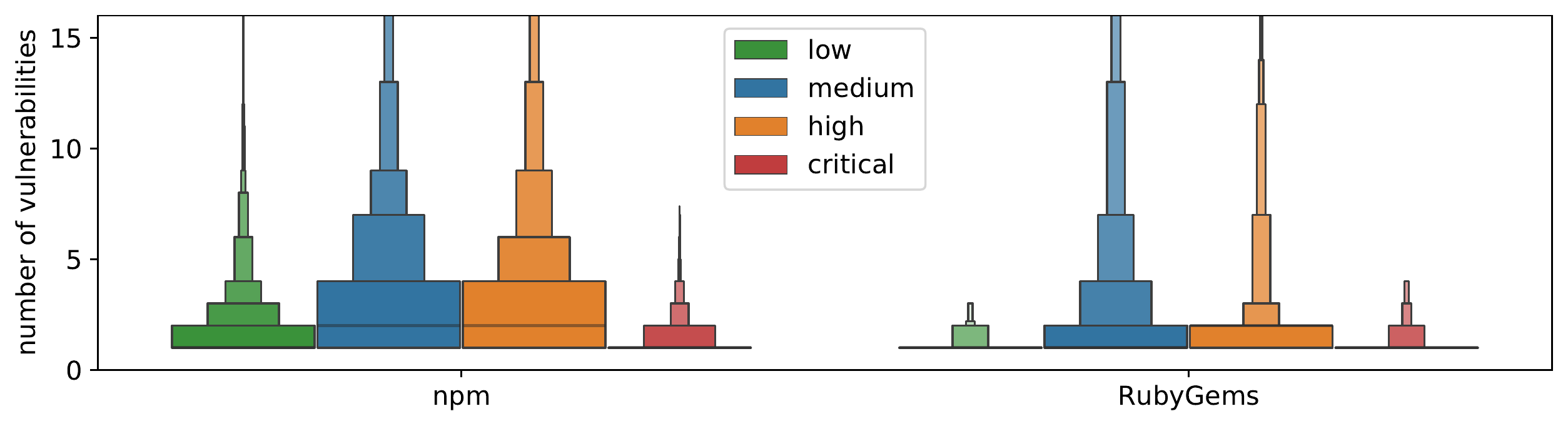}
		\caption{Boxen plots showing the distribution of the number of vulnerabilities in vulnerable {\bf indirect} dependencies of the \npm and \rubygems snapshots, grouped by severity.}
		\label{fig:transitive_deps}
	\end{center}
\end{figure}

Releases may also be exposed \emph{indirectly} to vulnerable dependencies.
1,225,724 out of the 64,959,052 indirect dependencies in \npm (1.9\%) are vulnerable;
whereas a much higher proportion of 65,090 out of 1,033,870 indirect dependencies in \rubygems (6.3\%) are vulnerable.
\fig{fig:transitive_deps} shows the distribution, per severity category, of the number of vulnerable package releases found among the indirect dependencies for \npm and \rubygems. Similar to \fig{fig:direct_deps}, medium and high severity vulnerabilities are the most common. We also observe a higher number of vulnerabilities for each severity level for \npm. %
Mann-Whitney U tests comparing the distributions between \npm and \rubygems confirm a statistically significant difference.
The effect sizes for low, medium, high and critical vulnerabilities are \textit{small} to \textit{moderate} in favor of \npm (see \tab{tab:rq3_severities_indirect}). Disregarding the severity category, the effect size is small (\ie $|d|=0.32$).

\begin{table}[!ht]
	\centering
	\caption{Mean and median number of vulnerabilities found in indirect dependencies, in addition to effect sizes and their directions when comparing \npm and \rubygems dependency vulnerabilities.}
\label{tab:rq3_severities_indirect}
\begin{tabular}{l|rr|rr||ccr}
	\multirow{2}{*}{} & \multicolumn{2}{c|}{\bf\npm} & \multicolumn{2}{c||}{\bf\rubygems} & \multirow{2}{*}{\bf direction} & \multirow{2}{*}{ $|\textbf{d}|$} & \multirow{2}{*}{\bf effect size}\\
	& \bf mean       & \bf median      & \bf mean         & \bf median         &                            &                              \\ \hline
	\bf \low              & 1.95       & 1           & 1.08         & 1              & \textgreater{}                & 0.43    & {\bf moderate}                     \\
	\bf \medium            & 3.53       & 2           & 2.77        & 1              & \textgreater{}                & 0.32               & small          \\
	\bf \high              & 3.41       & 2           & 2         & 2              & \textgreater{}             & 0.27            & small              \\
	\bf \critical          & 1.33       & 1           & 1.12         & 1              & \textgreater{}                & 0.15           & small               \\ \hline
	\bf all               & 3.08       & 2           & 1.94         & 1              & \textgreater{}             & 0.32 & small 
\end{tabular}
\end{table}

\smallskip
\noindent\fbox{%
	\parbox{0.98\textwidth}{%
 The proportion of vulnerable indirect dependencies for \rubygems (6.3\%) is more than three times higher than for \npm (1.9\%).
 On the other hand, for each severity, vulnerable indirect dependencies in \npm have more vulnerabilities than in \rubygems.
	  }%
}

\bigskip

\changed{
\paragraph{\textbf{Vulnerabilities of all transitive dependencies.}}

Considering both direct and indirect dependencies, we investigate whether the number of dependency vulnerabilities is related to the date when the studied package was released. \fig{fig:evolution_package_vulns} visualises the monthly evolution of the distribution of the number of dependency vulnerabilities for all packages released during that month. In general, we observe that for both ecosystems, the number of vulnerabilities decreased over time, in the sense that more recent packages are exposed to less vulnerabilities coming from their dependencies than older packages. However, we  notice that the number of vulnerabilities for \npm was increasing over time until 2014 (\ie dashed line in red), after which it started decreasing. This coincides with the time when \npm introduced the permissive constraint \textit{caret} (\caret) as default constraint for \npm dependencies instead of the more restrictive constraint \textit{tilde} ($\sim$). Caret accepts new patch and minor releases to be installed while tilde only accepts new patches. This means that packages with permissive dependency constraints are exposed to less dependency vulnerabilities than those with restrictive constraints. Intuitively, permissive constraints accept wider ranges of releases than restrictive ones and thus they provide more opportunities to install dependencies with fixed vulnerabilities.

\begin{figure}[!ht]
	\begin{center}
		\setlength{\unitlength}{1pt}
		\footnotesize
		\includegraphics[width=0.98\columnwidth]{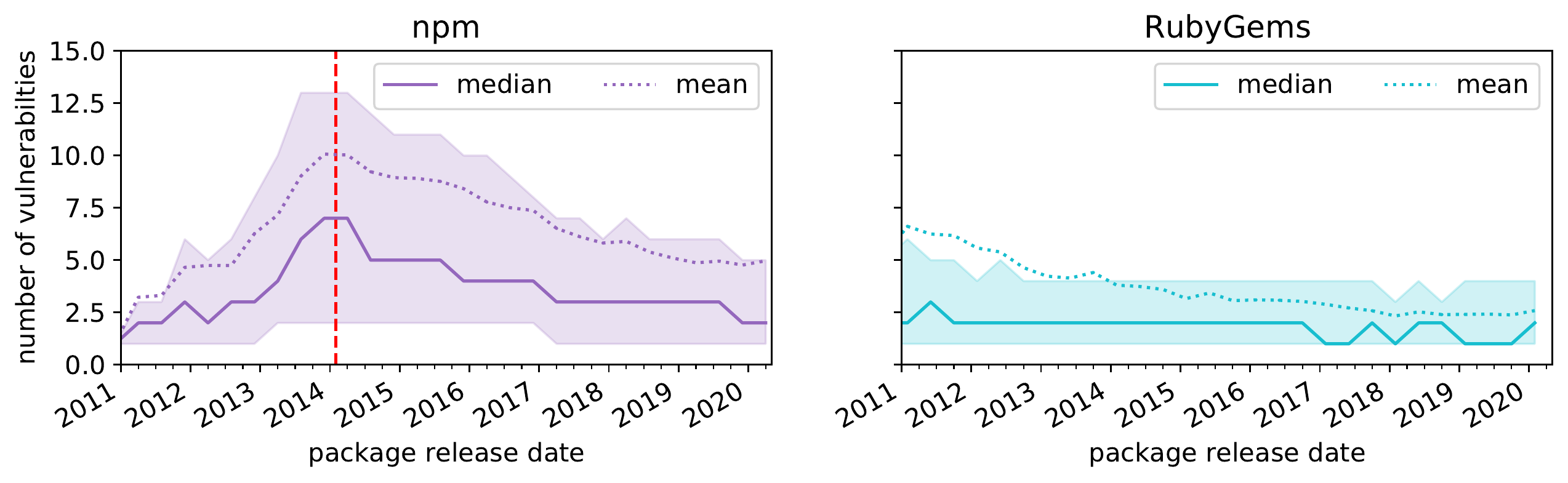}
		\caption{Monthly evolution of the distribution of the number of vulnerabilities coming from transitive dependencies of all studied packages. The shaded areas correspond to the interval between the $25^{th}$ and $75^{th}$ percentile.}
		\label{fig:evolution_package_vulns}
	\end{center}
\end{figure}

\smallskip
\noindent\fbox{%
	\parbox{0.98\textwidth}{%
		Older packages are exposed to more vulnerabilities coming from their dependencies than recent ones. The introduction of the permissive dependency constraint caret in \npm led to have packages with less vulnerable transitive dependencies.
	}%
}

}

\paragraph{\textbf{Exposed packages.}}

Only 849 vulnerable packages used as dependencies (667 for \npm and 182 for \rubygems) are responsible for all exposed packages. This means that 60.1\% and 43.3\% of vulnerable packages in \npm and \rubygems is never used as a dependency,
as the number of vulnerable packages we originally found is 1,672 for \npm and 321 for \rubygems. %
\changed{Moreover, only a small subset of the used vulnerable packages is responsible for most of the vulnerabilities found among direct and indirect dependencies. In fact, 90\% of the vulnerabilities found in \npm dependencies come from 50 packages only. For \rubygems this subset is even smaller with only 20 packages responsible for 90\% of the vulnerabilities.}

\begin{figure}[!ht]
	\begin{center}
		\setlength{\unitlength}{1pt}
		\footnotesize
		\includegraphics[width=0.98\columnwidth]{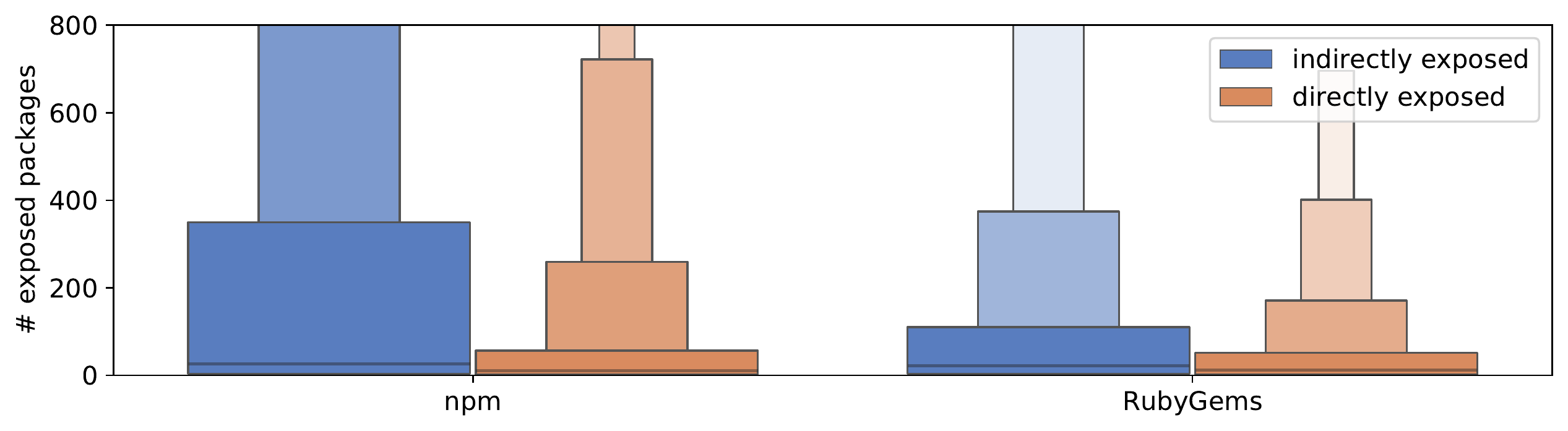}
		\caption{Distribution of the number of exposed \npm and \rubygems packages affected by a given vulnerable package.}
		\label{fig:exposedPackgesToDepsBoxen}
	\end{center}
\end{figure}

Starting from a unique vulnerable package that is used as a dependency, we quantify the number of dependent packages that are directly or indirectly exposed to a vulnerability because of it. \fig{fig:exposedPackgesToDepsBoxen} shows the distribution, revealing that considerably more packages are indirectly exposed to vulnerabilities. The median number of packages that is directly exposed to one vulnerable package is 11 for \npm and 12 for \rubygems, while it is about twice as high for indirect exposures (26 for \npm and 21.5 for \rubygems).
We carried out Mann-Whitney U tests to confirm that the distribution for indirectly exposed packages is higher than for directly exposed packages between the distributions. The null hypothesis could only be rejected for \npm with a \textit{small} effect size ($|d|=0.19$).
Without distinguishing between direct or indirect dependencies, one single vulnerable package is responsible for exposing a median of 21 and a maximum of 213,851 (67.8\%) \npm packages, and a median of 19 and a maximum of 22,233 (60.2\%) \rubygems packages, respectively.

\smallskip
\noindent\fbox{%
	\parbox{0.98\textwidth}{%
		A limited set of vulnerable packages is responsible for most of the vulnerabilities exposed through dependencies.
		One single vulnerable package can be responsible for exposing two thirds of all dependent \minor{latest} package releases.%
		
	}%
}

\subsubsection*{\textbf{\rqThreeB}}

The main purpose of package managers such as \npm and \rubygems is to facilitate depending on packages.
Therefore, not only packages but also external projects can be exposed to the vulnerabilities of their dependencies.
We intuitively expect external projects to be more exposed to vulnerabilities through their dependents than packages distributed through the package manager, as the maintainers of the latter intend their packages to be depended on.

For the 24,593 collected external projects (see \sect{subsec:dependency}) we searched for all dependencies %
referenced in the vulnerability dataset. 
We found 79\% external projects for \npm (11,003 out of 13,930) and 74.1\% for \rubygems (7,901 out of 10,663) with at least one direct or indirect vulnerable dependency. More specifically, for \npm 47\% of all external projects are directly exposed and 70.4\% are indirectly exposed; whereas for \rubygems 54\% of all external projects are directly exposed and 66.4\% are indirectly exposed~\footnote{The two categories of directly and indirectly exposed projects are non-exclusive.}.

\smallskip
\noindent\fbox{%
	\parbox{0.98\textwidth}{%
		\changed{
		About half of all external projects (47\% for \npm and 54\% for \rubygems) are exposed to vulnerabilities coming from vulnerable {\bf direct} dependencies. 
		About two thirds of all external projects (70.4\% for \npm  and 66.4\% for \rubygems) are exposed to vulnerabilities coming from vulnerable {\bf indirect} dependencies.}
	}%
}

\begin{figure}[!ht]
	\begin{center}
		\setlength{\unitlength}{1pt}
		\footnotesize
		\includegraphics[width=0.98\columnwidth]{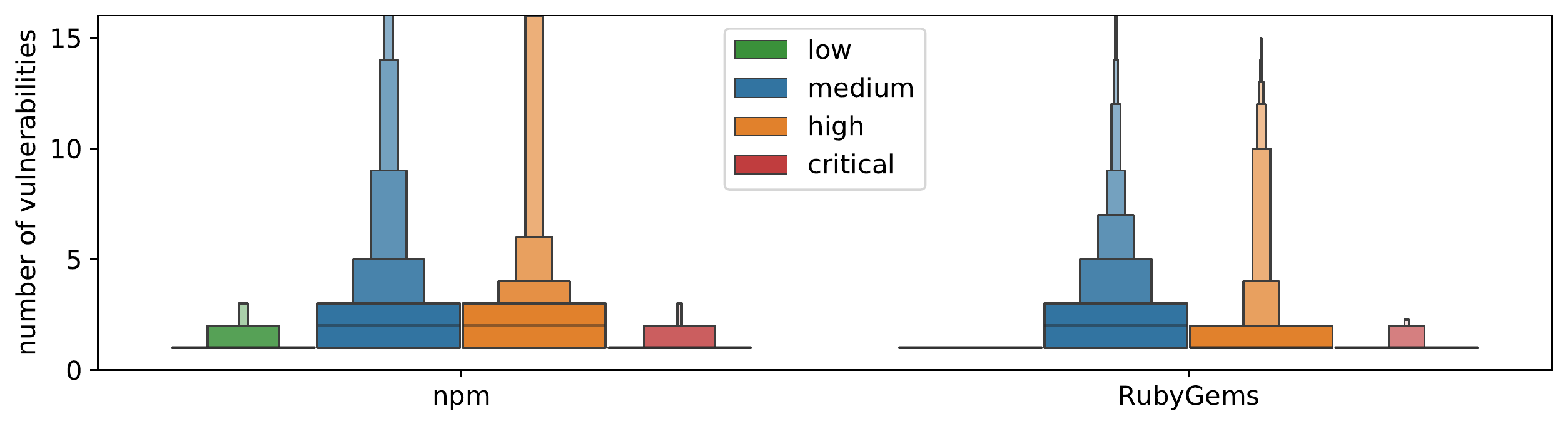}
		\caption{Boxen plots showing the distribution of the number of vulnerabilities found in vulnerable {\bf direct} \npm and \rubygems dependencies of \github projects, grouped by severity.}
		\label{fig:direct_repos}
	\end{center}
\end{figure}

\paragraph{\textbf{Vulnerable direct dependencies of external projects.}} 
Out of 147,622 of the direct dependencies of external projects on \npm packages, 11,969 (8.1\%) are vulnerable.
Out of 101,079 of the direct dependencies of external projects on \rubygems packages, 11,034 (10.9\%) are vulnerable.
\fig{fig:direct_repos} shows the distribution of the number of vulnerabilities affecting direct dependencies of external projects.
Similar to what we observed for $RQ_3^a$, medium and high severity vulnerabilities are the most common among direct dependencies, and dependencies on \npm packages tend to have higher numbers of vulnerabilities than dependencies on \rubygems packages.
\tab{tab:rq4_severities} shows the mean and median number of vulnerabilities caused by direct dependencies, grouped by severity.
We performed Mann-Whitney U tests to compare the number of vulnerabilities in project dependencies between \npm and \rubygems. 
We found statistically significant differences for all compared distributions but the effect size was \textit{negligible}, with one exception:
a \textit{moderate} effect size was found for highly vulnerable dependencies.

\begin{table}[!ht]
	\centering
	\caption{Mean and median number of vulnerable dependencies, in addition to effect sizes and their directions, for direct dependencies of \github projects on vulnerable \npm and \rubygems packages.}
	\label{tab:rq4_severities}
	\begin{tabular}{l|rr|rr||ccr}
		\multirow{2}{*}{} & \multicolumn{2}{c|}{\bf\npm} & \multicolumn{2}{c||}{\bf\rubygems} & \multirow{2}{*}{\bf direction} & \multirow{2}{*}{ $|\textbf{d}|$} & \multirow{2}{*}{\bf effect size}\\
		& \bf mean       & \bf median      & \bf mean         & \bf median         &                            &                              \\ \hline
		\bf\low               & 1.16       & 1          & 1.03         & 1              & \textgreater{}                & 0.11                & negligible         \\
		\bf\medium            & 2.96       & 2          & 2.80      & 2             & \textless{}                & 0.05              & negligible           \\
		\bf\high              & 2.80       & 2        & 1.78         & 1              & \textgreater{}             & 0.33              & {\bf moderate}           \\
		\bf\critical          & 1.17       & 1           & 1.07         & 1              & \textgreater{}                & 0.09           & negligible             \\ \hline
		\bf all               & 2.60       & 2          & 2.27        & 1              & \textgreater{}             & 0.06 & negligible
	\end{tabular}
\end{table}

\smallskip
\noindent\fbox{%
	\parbox{0.98\textwidth}{%
		8.1\% of the direct dependencies of external projects on \npm are vulnerable, while this is 10.9\% for \rubygems. 
		\npm-dependent projects have more highly vulnerable direct dependencies than \rubygems-dependent projects.
	}%
}

\begin{figure}[!ht]
	\begin{center}
		\setlength{\unitlength}{1pt}
		\footnotesize
		\includegraphics[width=0.98\columnwidth]{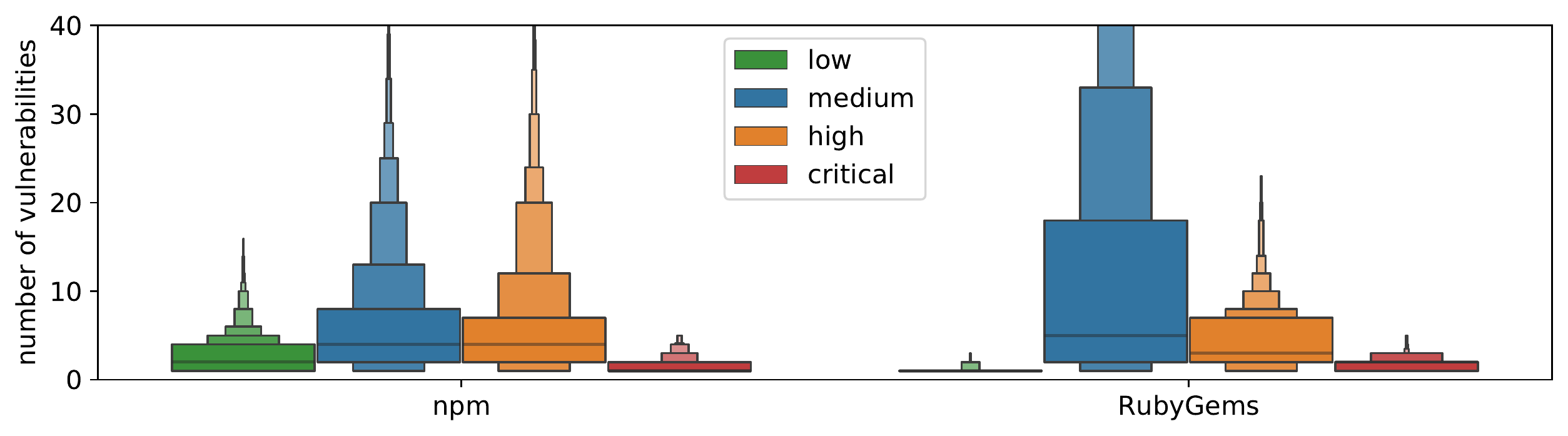}
		\caption{Distribution of the number of vulnerabilities found in vulnerable {\bf indirect} \npm and \rubygems dependencies of \github projects, grouped by severity.}
		\label{fig:transitive_repos}
	\end{center}
\end{figure}

\paragraph{\textbf{Vulnerable indirect dependencies of external projects.}} 
Out of 2,666,922 indirect dependencies of external projects on \npm packages, 87,062 (3.2\%) are vulnerable.
Out of 443,427 indirect dependencies of external projects on \rubygems packages, 46,682 (10.5\%) are vulnerable.
\fig{fig:transitive_repos} shows the distribution of the number of vulnerabilities affecting indirect dependencies of external projects.
Similar to \fig{fig:direct_repos}, medium and high severity vulnerabilities are the most common.
We observe that indirect dependencies on \rubygems tend to be more often of medium severity than indirect dependencies on \npm packages, while the latter tend be more often of low severity. 
Mann-Whitney U tests comparing the distributions between \npm and \rubygems confirmed a statistically significant difference.
Regardless of the severity level we found a \textit{small} effect size ($|d|=0.15$) in favour of \npm dependency vulnerabilities (see \tab{tab:rq4_severities_indirect}).
Per severity levels, we only found a non-negligible effect size for \textit{critical} severity vulnerabilities in favour of \rubygems ($|d|=0.25$) (i.e., \rubygems projects have more \textit{critical} vulnerabilities coming form indirect dependencies than \npm projects), and for \textit{low} severity vulnerabilities in favour of \npm ($|d|=0.71$). The \textit{negligible} effect size for \textit{medium} severity vulnerabilities is in favour of \rubygems ($|d|=0.1$), while for \textit{high} severity vulnerabilities it is in favour of \npm ($|d|=0.09$).

\begin{table}[!ht]
	\centering
	\caption{Mean and median number of vulnerable dependencies, in addition to effect sizes and their directions, for indirect dependencies of \github projects on vulnerable \npm and \rubygems  packages.}
	\label{tab:rq4_severities_indirect}
	\begin{tabular}{l|rr|rr||ccr}
		\multirow{2}{*}{} & \multicolumn{2}{c|}{\bf\npm} & \multicolumn{2}{c||}{\bf\rubygems} & \multirow{2}{*}{\bf direction} & \multirow{2}{*}{ $|\textbf{d}|$} & \multirow{2}{*}{\bf effect size}\\
		& \bf mean       & \bf median      & \bf mean         & \bf median         &                            &                              \\ \hline
		\bf \low              & 2.88       & 2           & 1.05         & 1              & \textgreater{}                & 0.71    & {\bf large}                     \\
		\bf \medium            & 6.56       & 4           & 12.36        & 5              & \textless{}                & 0.1               & negligible          \\
		\bf \high              & 6.2       & 4           & 4.55         & 3              & \textgreater{}             & 0.09            & negligible              \\
		\bf \critical          & 1.4       & 1           & 1.67         & 2              & \textless{}                & 0.25           & small               \\ \hline
		\bf all               & 5.16       & 3           & 5.74         & 2              & \textgreater{}             & 0.15 & small 
	\end{tabular}
\end{table}

\smallskip
\noindent\fbox{%
	\parbox{0.98\textwidth}{%
	Only 3.2\% of the indirect \npm dependencies of external projects are vulnerable, while this is more than three times higher (10.5\%) for \rubygems.
	Disregarding severities, external projects for \npm have more vulnerabilities coming from vulnerable indirect dependencies than for \rubygems.
		\rubygems external projects have more critical vulnerabilities coming from vulnerable indirect dependencies than \npm external projects, while the latter have more low severity vulnerabilities.%
	}%
}

\bigskip

\changed{
\paragraph{\textbf{Vulnerabilities of all transitive dependencies.}} 

\fig{fig:evolution_projects_vulns} visualises the monthly evolution of the distribution of the number of dependency vulnerabilities for all external projects that have their last commit during that month. Similar to \fig{fig:evolution_package_vulns}, we observe that more recently active projects are exposed to less vulnerabilities coming from their dependencies. Also, we can again see the impact of the introduction of the caret (\caret) dependency constraint in \npm in 2014 (left figure).	

	\begin{figure}[!ht]
		\begin{center}
			\setlength{\unitlength}{1pt}
			\footnotesize
			\includegraphics[width=0.98\columnwidth]{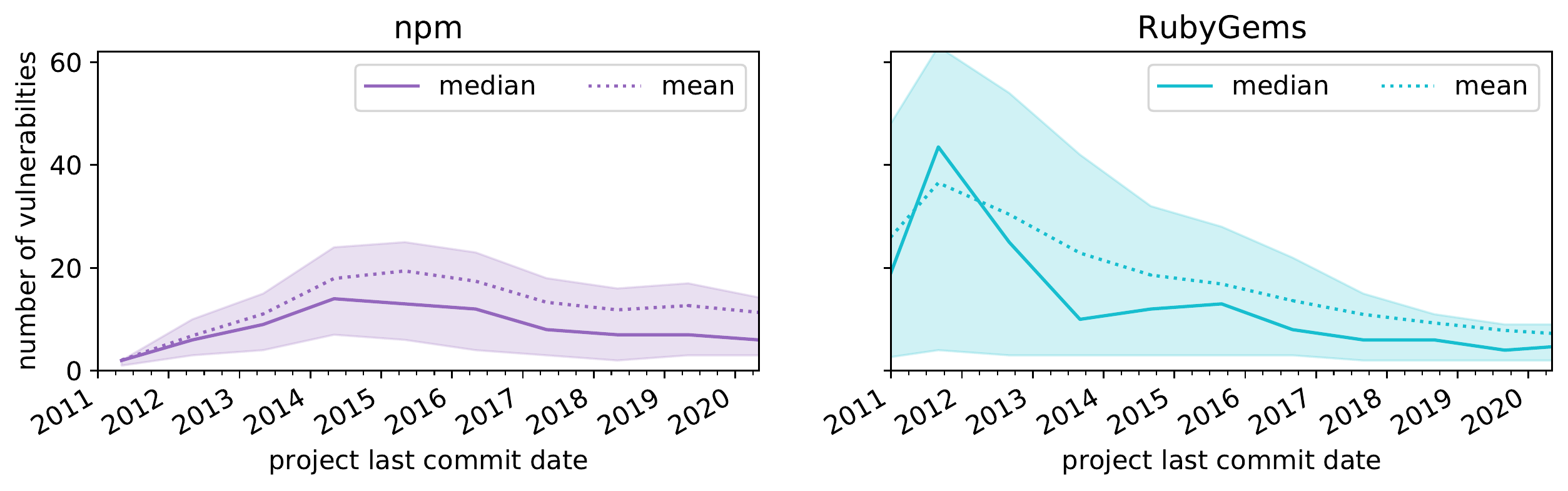}
			\caption{Monthly evolution of the distribution of the number of vulnerabilities coming from transitive dependencies of all studied external projects. \minor{Time points refer to the project's last commit date, \ie each project is considered only once.} The shaded areas correspond to the interval between the $25^{th}$ and $75^{th}$ percentile.}
			\label{fig:evolution_projects_vulns}
		\end{center}
	\end{figure}
	
	\smallskip
	\noindent\fbox{%
		\parbox{0.98\textwidth}{%
			More recently active external projects are exposed to \minor{fewer} vulnerabilities coming from their dependencies than older ones. The introduction of the permissive dependency constraint caret in \npm seems to have led to less vulnerable dependencies in external projects.
		}%
	}
}

\paragraph{\textbf{Exposed external projects.}} 
Only 560 (28\%) vulnerable packages (400 for \npm and 160 for \rubygems) are responsible for all exposed external projects.
This is less than the number we found for exposed packages in $RQ_3^a$, most likely because we studied fewer external projects than (internal) packages. 

\begin{figure}[!ht]
	\begin{center}
		\setlength{\unitlength}{1pt}
		\footnotesize
		\includegraphics[width=0.98\columnwidth]{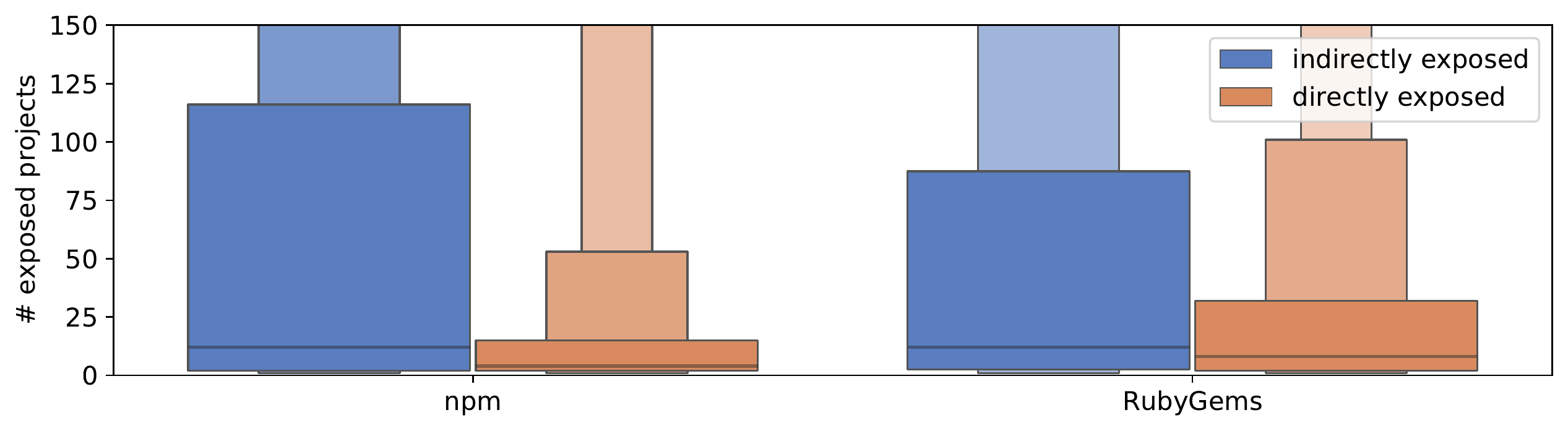}
		\caption{Distribution of the number of exposed \npm and \rubygems projects that one single vulnerable package is affecting.}
		\label{fig:exposedProjectsToDepsBoxen}
	\end{center}
\end{figure}

Similar to \fig{fig:exposedPackgesToDepsBoxen}, 
we analysed how many external projects are exposed to vulnerabilities because of a single vulnerable package. \fig{fig:exposedProjectsToDepsBoxen} shows the distribution of the number of projects that are directly or indirectly exposed to one unique vulnerable package. 
We observe that there are considerably more indirectly exposed projects than direct ones.
The median number of projects that one vulnerable \npm package directly exposes is 4, while for indirect exposure it is 12.
The median number of projects that one vulnerable \rubygems package directly exposes is 8, while for indirect exposure it is 12.

We carried out Mann-Whitney U tests between the distributions of direct and indirect dependency vulnerabilities exposing external projects. The null hypothesis could only be rejected 
in the case of \npm comparison with a \textit{small} effect size ($|d|=0.27$) in favor of indirectly exposed projects. 
Without distinguishing between direct or indirect dependencies, we found that one single vulnerable package is responsible for exposing a median of 8 and a maximum of 7,506 (68.2\%) projects on vulnerable \npm dependencies; while a median of 13.5 and a maximum of 5,270 (49.4\%) projects is exposed to vulnerable \rubygems dependencies.

\noindent\fbox{%
	\parbox{0.98\textwidth}{%
		Only 28\% of all vulnerable packages \minor{are} responsible for all vulnerabilities of exposed external projects.
		\changed{One single vulnerable package can be responsible for exposing 68.2\% of all exposed projects that use \npm, and 49.4\% of all exposed projects that use \rubygems.}

	}%
}
\subsection{\rqFour}
\label{subsec:rq4}
$RQ_3$ revealed that dependents are considerably more often indirectly exposed to vulnerabilities than directly. %
With $RQ_4$, we want to know how deep in the dependency tree we can find vulnerabilities to which packages and external projects are exposed. This allows us to quantify the transitive impact that vulnerable packages may have on their (transitive) dependents. 
In an earlier study focusing on the package dependency networks of 7 different package distributions (including \npm and \rubygems), \changed{Decan \etal~\cite{Decan2019}} analysed the prevalence of indirect dependencies. They observed that  more than 50\% of the top-level packages\footnote{Top-level packages are packages that do not have any dependent packages themselves.} in \npm have a dependency tree depth of at least 5, whereas for the large majority of top-level packages in \rubygems this was 3 or less. As a result, \npm appears to be potentially much more subject to deep vulnerable dependencies. $RQ_4$ aims to quantify this claim, \changed{focusing on \minor{the latest} package releases in $RQ_4^a$ and external projects in $RQ_4^b$.}

\subsubsection*{\textbf{\rqFourA}}

\begin{figure}[!ht]
	\begin{center}
		\setlength{\unitlength}{1pt}
		\footnotesize
		\includegraphics[width=0.98\columnwidth]{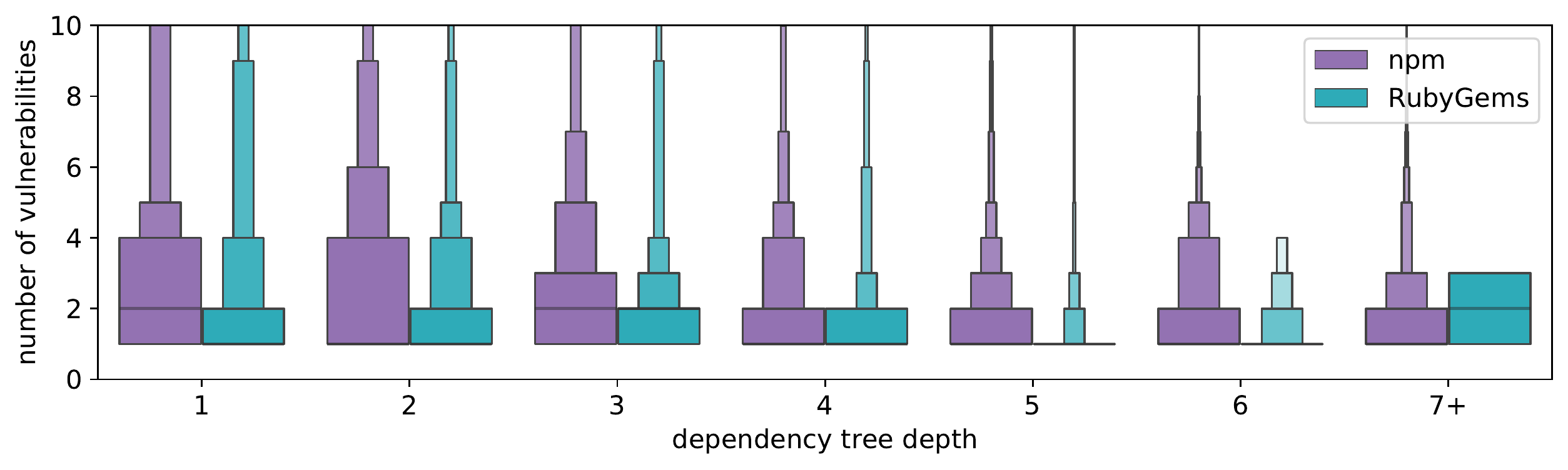}
		\caption{Distribution of the number of vulnerabilities found in all (direct and indirect) dependencies of \npm and \rubygems latest package releases, grouped by dependency tree depth.}
		\label{fig:propgation_packages}
	\end{center}
\end{figure}

We computed the number of vulnerabilities at each depth for all vulnerable (direct or indirect) dependencies. \fig{fig:propgation_packages} shows the distribution of the number of vulnerabilities in dependencies, grouped by dependency depth.
The first level (corresponding to direct dependencies) and the second level (dependencies of dependencies) have higher numbers of vulnerabilities. The number of vulnerabilities decreases at deeper levels, as a consequence of the fact that there are fewer releases with deep dependency trees.
Statistical comparisons confirmed, with non-negligible effect sizes, that more shallow levels correspond to higher numbers of vulnerabilities.
Nevertheless, vulnerabilities do remain present at the deepest levels. For example, we could find at least one vulnerable dependency at depth 16 for \npm package \textsf{formcore} that is indirectly exposed to a vulnerability in package \textsf{kind-of}, and at depth 10 for \rubygems package  \textsf{erp\_inventory} that is indirectly exposed to a vulnerability in package \textsf{rack}.

The vulnerable packages used as dependencies in \npm with the highest number of vulnerabilities 
are \textsf{node-sass}, \textsf{lodash} and \textsf{minimist}, while for \rubygems they are \textsf{nokogirl}, \textsf{activerecord} and \textsf{actionpack}. Unsurprisingly, we found the same set of vulnerable dependencies and vulnerability types reoccurring at deeper dependency levels.
The most prevalent vulnerability type for \npm was \emph{Prototype Pollution (PP)}, while for \rubygems dependencies it was \emph{Denial of Service (DoS)} (see \tab{tab:vuln_names} for an overview of the most common vulnerability types). 

\smallskip
\noindent\fbox{%
	\parbox{0.98\textwidth}{%
		The number of dependency vulnerabilities \changed{for the latest package releases} decreases at deeper levels of the dependency tree. Yet, vulnerable dependencies continue to be found at the deepest levels. The same vulnerability types can be found at all dependency tree levels. 
	}%
}

\begin{figure}[!ht]
	\begin{center}
		\setlength{\unitlength}{1pt}
		\footnotesize
		\includegraphics[width=0.98\columnwidth]{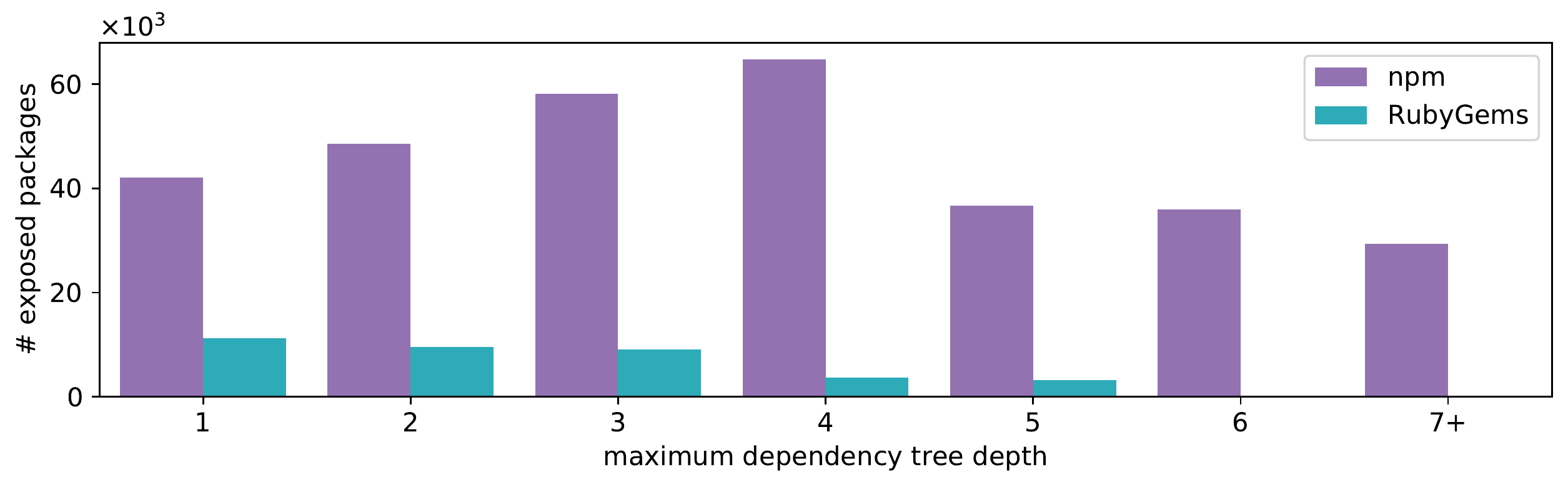}
		\caption{Number of exposed \npm and RubyGems \minor{latest} package releases and their maximum dependency depth.}
		\label{fig:numberOfExposedPackgesMax}
	\end{center}
\end{figure}

\smallskip
We also studied how deep in the dependency tree a vulnerability can reach by identifying for each exposed package the maximum dependency depth where a vulnerable dependency could be found (\fig{fig:numberOfExposedPackgesMax}). %
For \npm, the number of exposed packages 
increases from the dependency depth 1 (direct dependencies) to depth 4 and then starts to decrease. For \rubygems, the numbers start to decrease starting from depth 1. This implies that \npm packages are more susceptible to indirect vulnerable dependencies at deeper dependency levels. \changed{This seems to be in line with the findings of Decan \etal~\cite{Decan2019} mentioned at the beginning of this research question.}

\smallskip
\noindent\fbox{%
	\parbox{0.98\textwidth}{%
		\npm packages are more likely than \rubygems packages to be exposed to vulnerabilities deep in their dependency tree.	
	}%
}

\subsubsection*{\textbf{\rqFourB}}

\begin{figure}[!ht]
	\begin{center}
		\setlength{\unitlength}{1pt}
		\footnotesize
		\includegraphics[width=0.98\columnwidth]{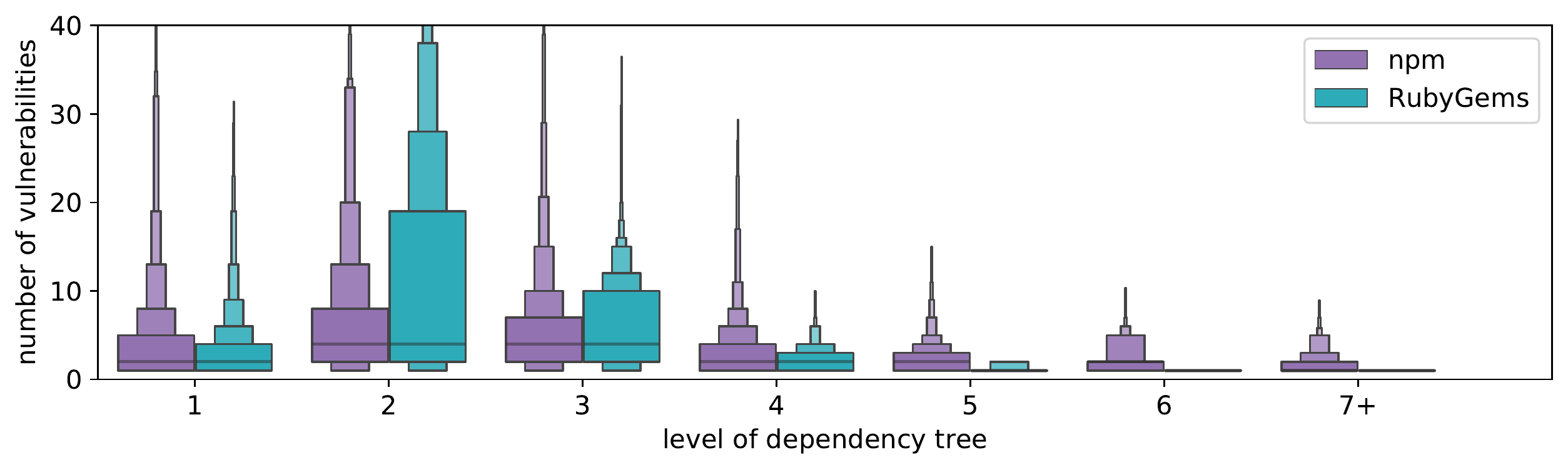}
		\caption{Distribution of the number of vulnerabilities found in all (direct and indirect) dependencies of \npm and \rubygems \github projects, grouped by dependency depth.}
		\label{fig:propgation_repos}
	\end{center}
\end{figure}

To gain more insights about the prevalence of vulnerabilities at different depths in the dependency tree of external projects depending on \npm or \rubygems packages, we computed the number of vulnerabilities found at each dependency depth.
\fig{fig:propgation_repos} shows the distribution of the number of vulnerabilities in dependencies, grouped by dependency depth. 
From the second level onwards, the number of vulnerabilities starts to decrease with increasing levels of depth.
As for $RQ_4^a$, statistical comparisons confirmed, with non-negligible effect sizes, that more shallow levels correspond to higher numbers of vulnerabilities.
Yet, vulnerabilities continue to remain present at the deepest levels.
For example, we could find at least one external project with a vulnerable dependency at depth 14 on \npm package \textsf{kind-of}, and one external project with a vulnerable dependency at depth 7 on \rubygems package \textsf{nokogiri}.
We also observe that external projects for \rubygems and \npm have the same median number of dependency vulnerabilities at depths 2 and 3, while starting from level 4 we observe the same trend as in \fig{fig:propgation_packages}. The reason for this difference at shallow depths (compared to what we saw in $RQ_4^a$) is because external projects depending on \rubygems tend to include more direct dependencies than normal \rubygems packages. 

\smallskip
\noindent\fbox{%
	\parbox{0.98\textwidth}{%
	Starting from depth 2, the number of vulnerable dependencies \changed{for external projects} decreases as the depth within the dependency tree increases.
	Yet, vulnerable dependencies are still found at the deepest levels.
	}%
}
\smallskip

\begin{figure}[!ht]
	\begin{center}
		\setlength{\unitlength}{1pt}
		\footnotesize
		\includegraphics[width=0.98\columnwidth]{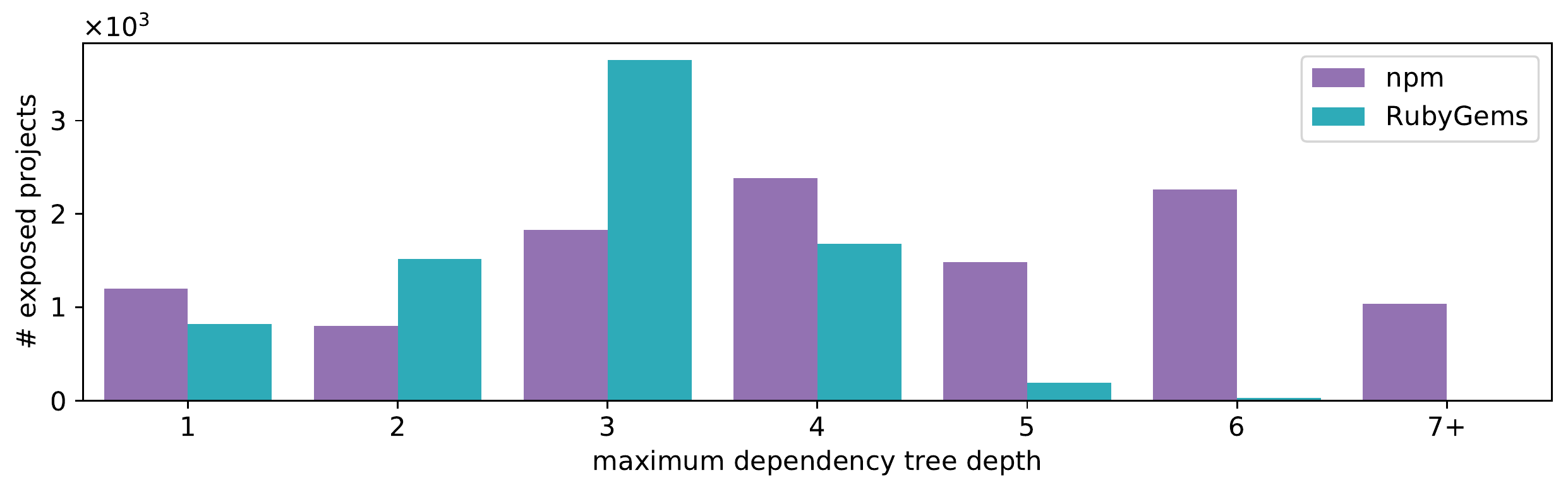}
		\caption{Number of exposed \changed{external projects for \npm and \rubygems} and their maximum dependency depth, grouped by dependency tree depth.}
		\label{fig:numberOfExposedProjectsMax}
	\end{center}
\end{figure}

We also analysed how far in the dependency tree a vulnerability can reach. \fig{fig:numberOfExposedProjectsMax} shows the number of exposed \changed{external} projects with the maximum depth at which we found at least one vulnerable dependency. We observe that for both \npm and \rubygems, the number of exposed \changed{external} projects increases from the first level (direct dependencies) until the fourth and third levels, respectively, and then starts decreasing.
 Comparing this finding with \fig{fig:numberOfExposedPackgesMax} in $RQ_4^a$, we observe that \rubygems projects are more susceptible to having vulnerable dependencies at deeper levels than \rubygems packages.

\noindent\fbox{%
	\parbox{0.98\textwidth}{%
		External projects dependent on \rubygems are more likely to be exposed to vulnerabilities deep in their dependency tree than external projects for \npm.	
	}%
}

\subsection{\rqFive}
\label{subsec:rq5}

57.8\% of the vulnerabilities in our dataset (\ie 1,611 to be precise) have a known fix.
With $RQ_5$ we aim to quantify how much  dependent packages and dependent external projects would benefit from updating their dependencies for which there is a known fix available. Upgrading their dependencies to more recent releases will reduce their exposure to vulnerabilities.
First, we start by exploring how many vulnerable dependencies have known vulnerability fixes. \tab{tab:rq5_fixes} shows the proportion of direct and indirect dependencies that are only affected by vulnerabilities that have a known fix. We observe that for the large majority of the vulnerable dependencies, fixes are available (more than 90\% for \npm and more than 60\% for \rubygems). \npm indirect dependencies have more fixed vulnerabilities than direct ones, while for \rubygems we observe the inverse. %
We also observe that fewer affected \rubygems dependencies have fixes available than \npm dependencies. These results show that most of the vulnerable dependencies could be made safe if the maintainers of the dependent packages or projects would choose the appropriate non-vulnerable version of their dependencies.

\begin{table}[!ht]
	\centering
	\caption{Proportion of vulnerable dependencies (for packages and external projects) having a known fix.}
	\label{tab:rq5_fixes}
	\begin{tabular}{l|r|r||r|r}

		\multirow{2}{*}{} & \multicolumn{2}{c||}{packages} & \multicolumn{2}{c}{projects} \\ \cline{2-5}
		& direct      & indirect      & direct             & indirect            \\ \hline
		npm               & 90.0          & 96.9            & 92.5               & 97.1                    \\
		RubyGems          & 85.2        & 66.7             &  90.0              & 76.0                  \\
	\end{tabular}
\end{table}

Updating a dependency is not always easy, especially when maintainers of dependent projects or packages would be confronted with breaking  changes.
 For example, if a \changed{package release or external project} uses a dependency that resolves to version 1.1.0 while a vulnerability fix exists in 1.2.0 then the exposure to the vulnerability can be removed by only doing a minor version update of the dependency. On the other hand, if a \changed{package release or external project} uses a dependency that resolves to version 1.1.0 while a vulnerability fix exists in 2.0.0, then one would need
to update to a new major version. If the dependency is adhering to \semver, then the  second but not the first kind of dependency update would require updating the dependent's implementation according to backwards incompatible changes.
We therefore verified whether it would be possible for dependent \changed{\minor{latest} package releases and external projects} to avoid vulnerabilities by only updating their vulnerable dependencies to a higher version within the same major version range that is currently in use.
Focusing only on direct dependencies, 32.8\% of the vulnerabilities affecting direct dependencies of \minor{latest} package releases and 40.9\% of the vulnerabilities affecting direct dependencies of external projects, could be avoided by making backward compatible dependency updates. For indirect dependencies, 22.1\% of the vulnerabilities affecting indirect dependencies of \minor{latest} package releases and 50.3\% of the vulnerabilities affecting indirect dependencies of external projects, have a fix within the major version range that is currently in use.
We also found that 5.4\% of the exposed \minor{latest} package releases and 5\% of the exposed external projects could be made completely vulnerability-free by only making backward compatible dependency updates to their vulnerable direct dependencies.

\smallskip
\noindent\fbox{%
	\parbox{0.98\textwidth}{%
	Vulnerability fixes are available for the large majority of vulnerable dependencies.
	Around one out of three dependency vulnerabilities to which \changed{the \minor{latest} package releases or external projects} are exposed, could be avoided if software developers would update their direct dependencies to more recent releases within the same major release range. 
	Performing backward compatible updates to vulnerable direct dependencies \changed{could} make 5.4\% of the exposed packages and 5\% of the exposed external projects completely vulnerability-free.
	}%
}
\minor{
\subsection{\rqSix}
\label{subsec:rq6}

$RQ_3$, $RQ_4$ and $RQ_5$ studied vulnerabilities of dependencies used in dependent packages and external projects as if they were deployed on 12 January 2020. This led us to consider all vulnerabilities already disclosed and reported in Snyk's dataset. $RQ_6$ investigates if dependents were already incorporating dependencies with disclosed vulnerabilities at their release time. To do so, we first need to resolve dependency constraints used in package releases and external projects at the time of their release. Then, we identify dependencies affected by disclosed vulnerabilities only. For example, the latest version of the package \texttt{node-sql}~\footnote{\url{https://www.npmjs.com/package/sql}} was released on August 2017 while depending on the package \texttt{lodash}~\footnote{\url{https://www.npmjs.com/package/lodash}}. If we resolve the used version of \texttt{lodash} on August 2017, we find that it was \texttt{4.1.0} which is affected by the vulnerability \texttt{CVE-2019-10744}~\footnote{\url{https://nvd.nist.gov/vuln/detail/cve-2019-10744}}. However, at the version release date in 2017, this vulnerability was not disclosed yet and thus the developers of \texttt{sql-node} could not do anything about it. Answering this research question will help us to assess how careful developers are when incorporating dependencies with already disclosed vulnerabilities. Therefore, we will only focus on direct dependencies.
\subsubsection*{\textbf{\rqSixA}}
For all 842,697 of the latest package releases available in the \npm and \rubygems snapshots, we determined their direct dependencies by resolving their dependency constraints at their release time. We narrowed down the analysis to those dependencies that are referenced in the vulnerability dataset.
6.8\% of all considered \npm latest package releases (50,720 out of 748,026) and 6.2\% of all considered \rubygems latest package releases (5,896 out of 94,671) were found to have, at their release dates, at least one direct dependency affected by at least one vulnerability that is already disclosed.

Moreover, of all 3,638,361 direct dependencies considered for \npm packages, only 58,184 (1.6\%) were affected by vulnerabilities disclosed before the latest package release in which the dependencies are incorporated. For \rubygems, of all 224,959 direct dependencies, only 6,410 (2.8\%) were affected. \tab{tab:rq6_severities} shows more details about the number of disclosed vulnerabilities found in direct dependencies. 

Comparing these results to $RQ_3^a$, we can clearly see that at their creation dates, the latest package releases were exposed to fewer disclosed vulnerabilities than on 12 January 2020 (the dataset snapshot date). Moreover, more than half of the package releases that are exposed to vulnerabilities via their dependencies at the snapshot date, were not exposed to any disclosed vulnerabilities when they were created. 

\begin{table}[!ht]
	\centering
	\caption{Mean and median number of disclosed vulnerabilities found in direct dependencies at the package release creation date, in addition to effect sizes and their directions when comparing \npm and \rubygems dependency vulnerabilities.}
	\label{tab:rq6_severities}
	\begin{tabular}{l|rr|rr|ccr}
		\multirow{2}{*}{} & \multicolumn{2}{c|}{\bf\npm} & \multicolumn{2}{c|}{\bf\rubygems} & \multirow{2}{*}{\bf direction} & \multirow{2}{*}{ $|\textbf{d}|$} & \multirow{2}{*}{\bf effect size}\\
		& \bf mean       & \bf median      & \bf mean         & \bf median         &                            &                              \\ 
		\toprule
		\bf \low              & 1.04       & 1           & 1.03         & 1              & \textgreater{}                & 0.01    & negligible                     \\
		\bf \medium            & 1.96       & 1           & 1.49         & 1              & \textgreater{}                & 0.1                & negligible          \\
		\bf \high              & 1.72       & 1           & 1.27         & 1              & \textgreater{}             & 0.2            & small              \\
		\bf \critical          & 1.07       & 1           & 1         & 1              & \textgreater{}                & 0.07           & negligible               \\ \hline
		\bf all               & 1.76       & 1           & 1.36         & 1              & \textgreater{}             & 0.1 & negligible 
	\end{tabular}
\end{table}

\smallskip
\noindent\fbox{%
	\parbox{0.98\textwidth}{%
		At their release dates, \rubygems latest package releases had proportionally more vulnerable direct dependencies than \npm (2.8\% compared to 1.6\%). 
		More than half of the latest package releases that are exposed to vulnerabilities via their dependencies at the observation date, were not exposed to any disclosed vulnerabilities when they were first created. 
	}%
}
\subsubsection*{\textbf{\rqSixB}}

For all 24,593 external projects that make use of \npm and \rubygems packages, we determined their direct dependencies by resolving their dependency constraints at their release time. 
22.1\% of all considered external projects for \npm (3,077 out of 13,930) and 33.9\% of all considered external projects for \rubygems (3,619 out of 10,663) were found to have, at their last commit date, at least one direct dependency affected by at least one vulnerability that is already disclosed. 

Out of 147,622 of the direct dependencies of external projects on \npm packages, only 4,600 (3.1\%) were affected by vulnerabilities disclosed before the date of the last commit. For \rubygems, of all 101,079 direct dependencies of external projects, only 5,264 (5.2\%) were affected. \tab{tab:rq6B_severities} shows more details about the number of disclosed vulnerabilities found in direct dependencies. 
\begin{table}[!ht]
	\centering
	\caption{Mean and median number of disclosed vulnerabilities found in direct dependencies of \github external projects at their last commit dates, in addition to effect sizes and their directions.}
	\label{tab:rq6B_severities}
	\begin{tabular}{l|rr|rr|ccr}
		\multirow{2}{*}{} & \multicolumn{2}{c|}{\bf\npm} & \multicolumn{2}{c|}{\bf\rubygems} & \multirow{2}{*}{\bf direction} & \multirow{2}{*}{ $|\textbf{d}|$} & \multirow{2}{*}{\bf effect size}\\
		& \bf mean       & \bf median      & \bf mean         & \bf median         &                            &                              \\ 
		\toprule
		\bf \low              & 1.16       & 1           & 1.02         & 1              & \textgreater{}                & 0.12    & negligible                     \\
		\bf \medium            & 2.29       & 1           & 1.65         & 1              & \textgreater{}                & 0.14                & negligible          \\
		\bf \high              & 1.88       & 1           & 1.38         & 1              & \textgreater{}             & 0.17            & small              \\
		\bf \critical          & 1.13       & 1           & 1.07         & 1              & \textgreater{}                & 0.04           & negligible               \\ \hline
		\bf all               & 2.01       & 1           & 1.55         & 1              & \textgreater{}             & 0.11 & negligible 
	\end{tabular}
\end{table}

Comparing these results to $RQ_3^b$, we observe that, at the time of their last commit, \github external projects were exposed to a lesser number of disclosed vulnerabilities than at the dataset snapshot date (\ie 12 January 2020). Moreover, more than half of the \github projects that make use of \npm packages and are exposed to vulnerabilities via their dependencies at the snapshot date, were not exposed to any disclosed vulnerabilities when they were first created. This is different for \github projects that make use of \rubygems packages, since only one third of the projects that are exposed to vulnerable dependencies at the snapshot date were not exposed to any vulnerability at the time of their last commit. 

\smallskip
\noindent\fbox{%
	\parbox{0.98\textwidth}{%
		At the time of their last commit, \github external projects that make use of \rubygems packages had proportionally more vulnerable direct dependencies than projects with \npm dependencies (33.9\% compared to 22.1\%). 
		Half of the external projects with \npm dependencies that are exposed to vulnerabilities at the observation date, were not exposed to any disclosed vulnerability at the time of their last commit, while this is only one third for projects with \rubygems dependencies.
	}%
}}
\section{Discussion}
\label{sec:discussion}

\changed{This section discusses our findings and their implications for developers, security researchers and package managers.
It provides insights specific to \npm and \rubygems, as well as some challenges that need to be overcome to better secure open source package distributions.}

We start our exposition with the vulnerable packages in each package distribution (\sect{sec:discuss-vuln-packages}), continue with the ramifications of how direct and indirect dependents are exposed to these vulnerabilities (\sect{sec:discuss-vuln-dependents}) \changed{and end with a discussion about comparing different package distributions (\sect{sec:discuss-comparing}).}

\subsection{Vulnerable packages}
\label{sec:discuss-vuln-packages}
In a comparative study of package distributions (for the observation period 2012--2017), Decan \etal~\cite{Decan2019} observed that both \npm and \rubygems have an exponential increase in their number of packages. The monthly number of package updates remained more or less stable for \rubygems, while a clear growth could be observed for \npm. 
$RQ_0$ builds further upon that work by analysing and comparing how vulnerable packages are in each package distribution using \snyk's dataset of vulnerability reports.

A first observation was that \npm has more vulnerabilities affecting more packages than \rubygems. We posit that this is due to the popularity of \npm that exposes the package distribution considerably more to attackers and attracts more security researchers. At the time of writing this article, the number of packages distributed through \npm is an order of magnitude higher than those distributed through \rubygems. \changed{In fact, we found that \npm has more security researchers that disclose vulnerabilities than \rubygems. Based on the adage ``Given enough eyeballs, all bugs are shallow"~\cite{maillart2017given}, the community of package repositories should invest in organizing bug bounty programs to attract more security researchers~\cite{alexopoulos2021vulnerability} that may help to discover hidden vulnerabilities. This will also help to reduce the vulnerability disclosure gap (see $RQ_1$).}

\changed{
\begin{leftbar}
	\vspace{-0.1cm}
	\noindent\textit{\textbf{Challenge:} How to incite more security researchers to inspect open source packages for vulnerabilities?}
\end{leftbar}
}

Let us now focus on vulnerability reports with the vulnerability type \textit{Malicious Package}. It is the most prevalent vulnerability type for \npm packages. %
Even though most packages suffering from this vulnerability will eventually be removed from the package distribution, previous studies~\cite{ohm2020backstabber} and experiments~\cite{dependencyConfusion} have shown that this type of vulnerability can be very dangerous.
\npm has 410 packages corresponding to the \textit{Malicious Package} vulnerability, while \rubygems has only a mere 19 vulnerabilities of this type.

The second most common vulnerability in \npm is \textit{Directory Traversal}, a vulnerability allowing an attacker to access arbitrary files and directories on the application server. This vulnerability could also be classified under two other types of attacks, namely \textit{Information Exposure} and \textit{Writing Arbitrary Files}. For \rubygems, we found \textit{Information Exposure} as the 4th ranked vulnerability type, probably because
its contributors prefer to report \textit{Directory Traversal} under the more general category of  \textit{Information Exposure} vulnerabilities.

\changed{
	\begin{leftbar}
		\vspace{-0.1cm}
		\noindent\textit{\textbf{Recommendation:} Malicious packages are more prevalent in \npm than \rubygems. \npm users should be aware of the common strategies of malicious packages like Typosquatting~\cite{ohm2020backstabber}.}
	\end{leftbar}
}

42\% (1,175) of the reported vulnerabilities in the dataset did not have any known fix. 
We can exclude from them 406 unfixed \textit{Malicious Package} vulnerabilities (389 for \npm and 17 for \rubygems)
that can ultimately be fixed by simply removing the package or the affected releases from the distribution.
This leaves us with 669 vulnerabilities different from the \textit{Malicious Package} type that affect all recent releases of \npm packages in which they were discovered, and 100 such vulnerabilities for \rubygems packages. \minor{Users of package distributions should be aware of whether package managers incorporate and mark the vulnerable packages and their releases in their registries.}
Having such information in package managers will help developers and users to decide which releases they should depend upon. It will also simplify the work of tools like \textsf{Dependabot}~\footnote{\url{https://dependabot.com/}} and \textsf{Up2Dep}~\cite{nguyen2020up2dep} that try to find and fix insecure code of used dependencies. Such tools should also be aware that not all vulnerabilities are in MITRE or NVD~\footnote{\url{https://nvd.nist.gov/}} since we found that 45\% of all vulnerabilities do not have a CVE reserved for them. Many of the vulnerability threats and their information are shared through social media channels (\eg Reddit, Twitter, Stack Overflow, etc)~\cite{aranovich2021beyond}. Therefore, other vulnerability databases, security advisories and issue trackers might help these tools to extend their reach to vulnerabilities that are not in NVD.

\changed{
	\begin{leftbar}
		\vspace{-0.1cm}
		\noindent\textit{\textbf{Recommendation:} Package managers can help developers by marking vulnerable package releases in their registries.}
	\end{leftbar}
}

\changed{\minor{To offer more secure packages, all known vulnerabilities should be fixed, discovered and disclosed rapidly.} Unfortunately, many vulnerabilities remain undisclosed for months to years, providing attackers plenty of time to search for flaws and develop exploits that can harm millions of users of these packages. 
For \npm, we observed that {\em critical} vulnerabilities were \changed{disclosed} more rapidly, while for \rubygems we did not observe a relation to the severity of the vulnerability. %
Software package maintainers should invest more effort in keeping their packages secure, especially those packages that are frequently used as dependents of other packages or external projects, in order to reduce the transitive exposure to vulnerabilities.
They should be supported in this task by additional static and dynamic application security testing tools, perhaps tailored to the characteristics of packages and libraries. 
For instance, there is typically no single method that can serve as the entry point for a whole-program analysis. 

Moreover, the longer a vulnerability lasts, and the more releases are affected by it, the more likely it becomes that someone will unknowingly use a vulnerable release. 
Registries of snapshot-based deployments such as \dockerhub that are attracting more and more popularity exacerbate this problem~\cite{zerouali2021multi,zerouali2021usage} as users might unknowingly reuse images that wrap vulnerable package releases. Unfortunately, the majority of the vulnerabilities in \npm and \rubygems packages are fixed within several years after their first introduction leaving most of the package releases affected. 
\minor{However, after their disclosure, vulnerabilities seem to be fixed after a short time. Only about one out of five vulnerabilities takes more than 3 months to be fixed, which is the deadline given by \snyk to maintainers before publishing a reported vulnerability publicly. This means that most of the maintainers are willing to fix vulnerabilities as soon as possible after disclosure.}
}

\changed{
	\begin{leftbar}
		\vspace{-0.1cm}
		\noindent\textit{\textbf{Challenge:} A vulnerability can stay hidden in a package for years and it affects most of its previous releases before it is finally fixed. Effective static and dynamic analysis tools should emerge to support security researchers in finding vulnerabilities.}
	\end{leftbar}
}

\subsection{Dependents exposed to vulnerabilities}
\label{sec:discuss-vuln-dependents}

$RQ_3^a$ investigated how packages are exposed to vulnerable dependencies. To this end, we studied the latest release of each package available in the considered snapshot of each package distribution. Since previous studies~\cite{Cox2015_freshness,zerouali2021multi}
showed that less recent releases of a package tend to be more vulnerable, we expect to find more vulnerable dependencies if we would study all package releases rather than only the latest one.
The results of $RQ_3^a$ revealed that only a small set of vulnerable packages %
is responsible for a large number of vulnerable dependencies. Moreover, $RQ_4^a$ showed that these vulnerable dependencies can often be found deep in the dependency tree, making it difficult for dependents to cope with them. \changed{Several very popular packages that are used by many dependents have been affected by vulnerabilities, leading to a vulnerability exposure in thousands of transitively dependent packages.}

\changed{
	\begin{leftbar}
		\vspace{-0.1cm}
		\noindent\textit{\textbf{Recommendation:} Package maintainers should inspect not only direct, but also indirect dependencies, since vulnerabilities are often found deep in the dependency chain.}
	\end{leftbar}
}

\changed{
	\begin{leftbar}
		\vspace{-0.1cm}
		\noindent\textit{\textbf{Challenge:} How can the community help to secure popular packages, that are present in many dependency chains?}
	\end{leftbar}
}

Another factor behind this vulnerability exposure 
is that most of the reported vulnerabilities do not have a lower bound on affected releases. They are found to affect all package releases before the version in which the vulnerability was \changed{fixed}, 
yielding a wide range of vulnerable package releases. For example, \minor{according to Snyk,} the critical vulnerability CVE-2019-10744 affects all releases before version 4.17.12 of the popular package {\sf lodash}~\footnote{\url{https://nvd.nist.gov/vuln/detail/cve-2019-10744}}. \minor{However, maintainers of \texttt{lodash} showed vigilance since this vulnerability was fixed exactly 13 days after its disclosure. Users of package distributions therefore have an important responsibility in keeping the ecosystem in a healthy shape by keeping their own dependencies up to date.}
However, managers of package distributions \minor{can share this} responsibility by raising awareness of the importance of keeping dependencies to popular packages up to date. %
Maintainers of those popular packages \minor{can participate as well by informing} their dependents whenever vulnerabilities are discovered and fixes become available in newer versions, and maintainers of dependents that should actually update their dependencies to those fixed releases. 

\changed{
	\begin{leftbar}
		\vspace{-0.1cm}
		 \noindent\textit{\textbf{Recommendation:} \minor{Package developers should be aware that disclosed vulnerabilities frequently affect all previous releases of a package. This kind of vulnerabilities can be prioritized since all dependents of the package will be impacted.}}
	\end{leftbar}
}

\changed{
	\begin{leftbar}
		\vspace{-0.1cm}
		\noindent\textit{\textbf{Challenge:} How often should one inspect previous releases for a vulnerability that is disclosed in a recent release, and how accurate is this information in vulnerability report databases?~\cite{dashevskyi2018screening, nguyen2016automatic, meneely2013patch}}
	\end{leftbar}
}

\smallskip
The analysis of $RQ_3^b$ revealed that \changed{external projects hosted on \github}, containing software that is not distributed via the package distributions, are exposed to high numbers of vulnerabilities coming from transitive dependencies caused by a small set of vulnerable packages. %
Package dependents should rely on tools like {\sf npm audit} to run security checks and be warned about such vulnerable dependencies. %
We also noticed that many \npm vulnerabilities are coming from duplicated dependencies across the dependency tree. %
Dependents should reduce the number of duplicated \npm dependencies --~especially vulnerable ones~-- by sharing the common dependencies using commands like {\sf npm dedupe}~\footnote{\url{https://docs.npmjs.com/cli/v7/commands/npm-dedupe}} available for \npm. \changed{In a similar vein, dependents should check for bloated vulnerabilities which are not necessary to build or run the dependent software and remove them. This may reduce the number of vulnerable dependencies and their vulnerabilities. In fact, previous studies about dependencies in Maven showed that 2.7\% and 57\% of direct and transitive dependencies of Maven libraries are bloated~\cite{soto2021comprehensive}.}

\changed{
	\begin{leftbar}
		\vspace{-0.1cm}
		\noindent\textit{\textbf{Recommendation:} \npm dependents should reduce the relatively large number of vulnerable indirect dependencies by	eliminating duplicate package releases.}
	\end{leftbar}
}

\changed{We found that about 40\% of the packages and 70\% of the external projects have at least one vulnerable transitive dependency. This does not necessarily mean that the dependents are at risk and affected by the vulnerabilities of their dependencies. Many of the vulnerabilities will only affect the functionalities that are not actually used by the dependent~\cite{pashchenko2020vuln4real}. When fixing vulnerabilities coming from dependencies, developers should prioritize vulnerabilities by identifying the ones that are actually exposing their software by distinguishing between effective and ineffective functionalities used from dependencies. However, there are many cases where it is not necessary to use the vulnerable dependency to be affected by its vulnerability, \eg Prototype Pollution~\footnote{\url{https://www.whitesourcesoftware.com/resources/blog/prototype-pollution-vulnerabilities/}} and Malicious Package vulnerabilities~\cite{dependencyConfusion}. For this reason, it is important to know about all vulnerabilities coming from both direct and indirect dependencies and then decide whether they are exposing the dependent software or not.

\changed{
	\begin{leftbar}
		\vspace{-0.1cm}
		\noindent\textit{\textbf{Recommendation:} Indirect dependencies come with a high number of vulnerabilities, especially in \npm. Dependents should reduce the number and depth of their indirect dependencies or monitor them alongside the direct ones. }
	\end{leftbar}
}

Verifying direct dependencies for effective functionalities could be done by the developers of software dependents, while this could be difficult to do with transitive dependencies. Developers of package and project dependents may rely on Software Composition Analysis (SCA) tools to check transitive dependencies for vulnerabilities affecting their used dependency functionalities. They can also combine SCA tools to identify and mitigate the maximum number of vulnerabilities. In fact, previous studies have shown that SCA tools vary in their vulnerability reporting~\cite{imtiaz2021comparative}.}

\changed{
	\begin{leftbar}
		\vspace{-0.1cm}
		\noindent\textit{\textbf{Recommendation:} Package dependents should rely on available tools to run security checks and be warned about vulnerable dependencies and their fixes. }
	\end{leftbar}
}

\changed{
	\begin{leftbar}
		\vspace{-0.1cm}
		\noindent\textit{\textbf{Challenge:} Should developers be warned about all \minor{disclosed} vulnerabilities of their dependencies or only about vulnerabilities affecting the functionalities they use?~\cite{Pashchenko2018,pashchenko2020vuln4real,Ponta2020EMSE}}.
	\end{leftbar}
}

$RQ_5$ highlighted that most of the vulnerabilities exposing dependent packages and projects have a known fix, implying that \changed{by updating their vulnerable dependencies, dependents might} avoid those vulnerabilities. To keep informed about new releases that may include vulnerability fixes, dependents can rely on tools like \textsf{Dependabot}, which creates pull requests to update outdated and vulnerable direct dependencies to newer and patched releases.
It is the responsibility of package maintainers to keep their own dependencies up to date so dependent packages and projects can have secure transitive dependencies.
To do so, package dependents can rely on permissive \semver constraints~\cite{decan2019package} when specifying the dependencies to rely on in manifests like \emph{package.json} and \emph{Gemfile}. This will ensure that dependencies get updated automatically. For project maintainers, it is possible to use commands like \texttt{npm update PACKAGE --depth= DEPTH} to update their transitive dependencies and then lock their dependencies using lockfiles like \emph{package-lock.json}~\footnote{\url{https://docs.npmjs.com/cli/v7/configuring-npm/package-lock-json}} and \emph{Gemfile.lock}~\footnote{\url{https://bundler.io/rationale.html}}. \changed{An other option could be to update frequently. There is a big difference between updating a dependency with a security patch with a change of few lines of code versus several years worth of code~\cite{zerouali2019measurement}.} The analysis of $RQ_5$ showed that around one in three vulnerabilities \changed{might} be avoided if every dependent updated its dependencies to a newer minor or patch increment of the major release it is already using. This means that many vulnerable dependencies can be fixed just by modifying the dependency constraints to accept new minor and patch releases. On the other hand, for vulnerable dependencies for which the fix requires a major update, maintainers of dependent packages and projects should be careful as major dependency updates might introduce breaking changes. Therefore, maintainers of required packages should not only provide fixes in new major releases but should also attempt to provide these fixes in older major releases via backports\changed{~\cite{decan2021back}}. This way, even dependents that need to stick to older major releases could still benefit from the backported vulnerability fix. \changed{Perhaps popular packages with a high package centrality~\cite{mujahid2021towards} can benefit from community support in bringing such backports to older major releases.}

\changed{
	\begin{leftbar}
		\vspace{-0.1cm}
		\noindent\textit{\textbf{Recommendation:} Packages and external \github projects should invest more efforts in updating their dependencies.}
	\end{leftbar}
}

\changed{In addition, in $RQ_2$ we found a considerable proportion of vulnerabilities that have been fixed in minor and major releases. We think that package maintainers should try as much as possible to fix their vulnerabilities in patch updates or at least in backward compatible releases and follow semantic versioning. If a new package release incorporates breaking changes, then the major version number should be incremented. This will help developers to know whether they can update their vulnerable dependencies or not. Previous studies showed that developers are hesitant to update their vulnerable dependencies because they are afraid that the new releases not only include security fixes but also bundle them with functional changes. This hinders adoption due to lack of resources to fix functional breaking changes~\cite{pashchenko2020qualitative,nguyen2020up2dep}.}

\changed{
	\begin{leftbar}
		\vspace{-0.1cm}
		\noindent\textit{\textbf{Recommendation:} Vulnerability fixing updates should not include breaking changes. If package maintainers can rely on semantic versioning, they will be able to characterize their package updates, while their dependents will be able to decide whether they want to update their dependency or not.  }
	\end{leftbar}
}

\changed{
	\begin{leftbar}
		\vspace{-0.1cm}
		\noindent\textit{\textbf{Challenge:} Is it always possible to incorporate vulnerability fixes within patch updates?}
	\end{leftbar}
}

\subsection{Comparing package distributions}
\label{sec:discuss-comparing}

\changed{Our analysis revealed many differences between \npm and \rubygems.}
\npm has more reported vulnerabilities than \rubygems, and \npm projects exposed to vulnerabilities have more dependencies. On the other hand, \rubygems has higher proportions of vulnerable dependencies than \npm. It is likely that the security efforts undertaken by \npm~--such as its integrated dependency auditing tools--~are gradually making \npm more secure. 
\changed{Still, tooling could be improved further, as most} of the existing tooling relies on dependency metadata alone, \ie available security monitoring tools only rely on dependency information extracted from manifests like \textit{package.json} and \textit{Gemfile} to detect vulnerabilities. 
Usage-based vulnerability detection tools~\cite{zimmermann2019small,Ponta2020EMSE} should emerge to help developers in identifying which dependency vulnerabilities can actually be exploited.

\changed{
	\begin{leftbar}
		\vspace{-0.1cm}
		\noindent\textit{\textbf{Lesson learned:} More effort is needed to better secure open source package distributions. All parties can help including package managers, security experts, communities and developers.}
	\end{leftbar}
}

\changed{However, presence or absence of tooling is by no means the only factor that influences the proportion of vulnerable packages and vulnerable dependencies in a packaging ecosystem. Decan \etal \cite{Decan2019} have shown important differences in the topological structure and evolution of package dependency networks, whereas Bogart \etal \cite{Bogart2021} have shown that each ecosystem uses different practices and policies of their communities. All these factors are likely to play a role in vulnerability management.}

While this paper focused on \npm and \rubygems, Alfadel~\etal\cite{alfadelempirical} studied the PyPI package distribution. 
Through an analysis of 550 vulnerability reports affecting 252 \python packages they studied the time to \changed{disclose} and fix vulnerabilities in the PyPI package distribution.
There are some similarities between their results and our own observations for \npm and \rubygems.
For example, the number of vulnerabilities found in \pypi increases over time and the majority of those vulnerabilities are of medium or high severity. The most prevalent vulnerabilities in \pypi are \emph{Cross-Site-Scripting (XSS)} and \emph{Denial of Service (DoS)}, which is similar to what we found for \rubygems. Vulnerabilities in \pypi are \changed{disclosed} after a median of 37 months, which is similar to what we found for \npm. 
On the other hand, vulnerabilities in \pypi seem to take longer to be fixed than those in \npm and \rubygems. %
Since Alfadel~\etal\cite{alfadelempirical} did not study the exposure of dependent packages to vulnerabilities, we cannot compare \pypi to \npm and \rubygems on this aspect.

\changed{
	\begin{leftbar}
		\vspace{-0.1cm}
		\noindent\textit{\textbf{Challenge:} What are the main factors in a packaging ecosystem that play a role in its security management?}
	\end{leftbar}
}

 %
 %

%
%
%
%

%
%
%
%
%
%\section{Threats to validity}
\label{sec:threats}

The empirical nature of our research exposes it to several threats to validity. We present them here, following the classification and recommendations of~\cite{Wohlin:2000}.

The main threat to \emph{construct validity} comes from imprecision or incompleteness of the data sources we used to identify vulnerabilities and their affected and exposed packages. \changed{We assumed that the Snyk vulnerability database represents a sound and complete list of vulnerability reports for third-party packages. This may have led to an underestimation since some vulnerabilities may not have been disclosed yet and are therefore missing from the database.} Another data source that we relied on is \librariesio. While there is no guarantee that this dataset is complete (\eg there may be missing package releases), we did not observe any missing data during a manual inspection of the dataset. Considering the full set of packages ever released also constitutes a threat to validity since some of them could have been removed from the package registry yet are still referenced in \librariesio. To mitigate this threat we sanitised the dataset by excluding a number of \npm packages that were removed from \npm. Examples are the {\sf wowdude-x}~\footnote{\url{https://libraries.io/search?q=wowdude}} and {\sf neat-x} packages~\footnote{\url{https://libraries.io/npm/neat-106}}. The only purpose of these packages was to bundle a large set of run-time dependencies. 
\minor{
As explained in \sect{subsec:vulnerability}, the vulnerability  severity labels are extracted from Snyk. This data source has its own way of computing the severity of vulnerabilities. Relying on another data source might have led to different  vulnerability severity results. To understand how different severity labels in Snyk compare to other sources, we extracted the severity labels for all vulnerabilities with a CVE from NVD~\footnote{\url{https://nvd.nist.gov/}}. For all 1,487 vulnerabilities (55\% of the entire dataset) that we found with a CVE ID in $RQ_1$, only  1,227 have a severity label in NVD. 792 (64.55\%) of them have the same severity in both Snyk and NVD. The heatmap of \fig{fig:heatmap_severity} shows the proportion of vulnerabilities with different severity labels in Snyk and NVD. We observe that 21.68\% of the vulnerabilities have higher severity in NVD than in Snyk, while 13.77\% of the vulnerabilities are more severe in Snyk than in NVD. This confirms that the findings could differ if another vulnerability data source were to be used. For example, the observed vulnerabilities exposing package and project dependents could be more severe than what is reported in this paper. Nevertheless, since many vulnerabilities are not even reported by NVD, we chose for the pragmatic option to rely on the more complete dataset of Snyk.

We also tried to verify severity labels given by CNAs (\ie vulnerability reporters in NVD), but only 42 vulnerabilities contained such labels.
	
	\begin{figure}[!ht]
		\begin{center}
			\setlength{\unitlength}{1pt}
			\footnotesize
			\includegraphics[width=0.9\columnwidth]{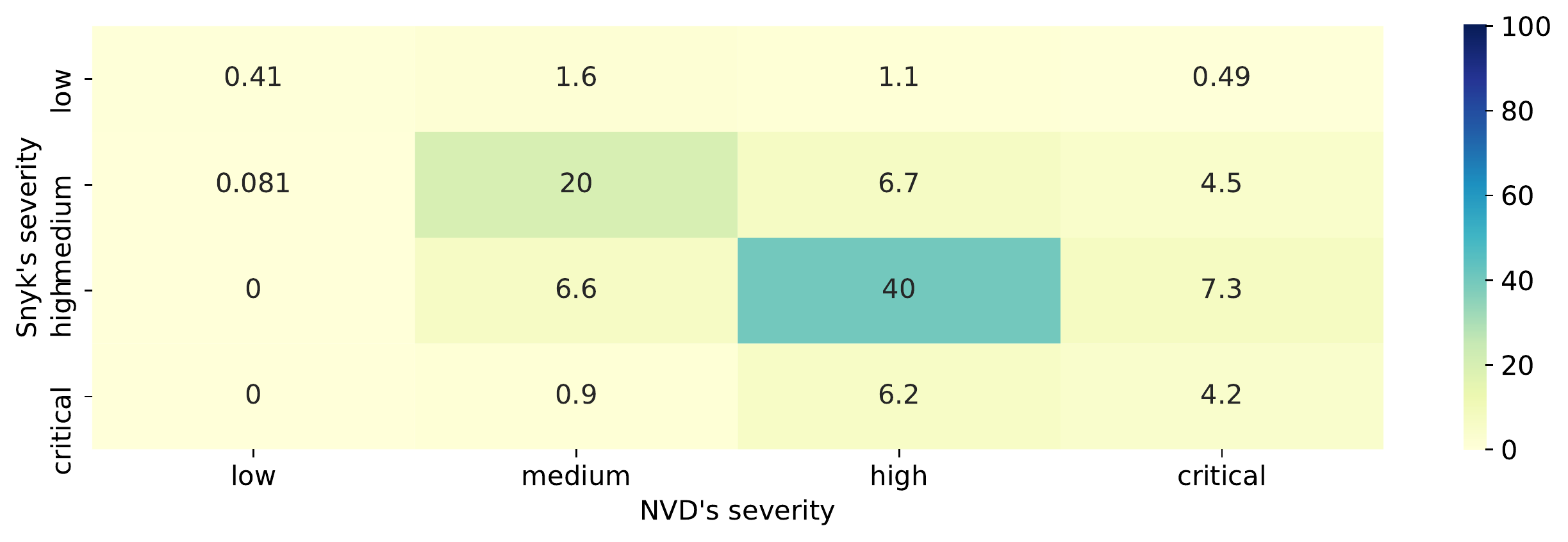}
			\caption{Heatmap of the proportion of vulnerabilities that have the same or different severity labels in Snyk and NVD.}
			\label{fig:heatmap_severity}
		\end{center}
	\end{figure}
}

Another threat to construct validity stems from the fact that we found some packages whose first vulnerable releases were never distributed via the package manager. %
For example, the cross-site scripting (XSS) vulnerability type~\footnote{\url{https://snyk.io/vuln/npm:wysihtml:20121229}} affects all releases before version 0.4.0 of the \npm package \texttt{wysihtml}. Those releases, including version 0.4.0, were never actually distributed through \npm~\footnote{\url{https://www.npmjs.com/package/wysihtml}} even though they are present on the package's \github repository~\footnote{\url{https://github.com/Voog/wysihtml/tags?after=0.4.0}}. Since our study only focused on vulnerabilities affecting package releases contained in the \librariesio dataset, our results might underestimate the exact time needed to \changed{disclose} a vulnerability.

We noticed that many of the reported vulnerabilities affect {\em all} package releases before the version in which the vulnerability was fixed, leading to a large number of vulnerable package releases. \changed{
To ascertain that indeed all package releases before the fix are affected,} we contacted the \snyk security team. They confirmed that they carry out manual inspections of the affected releases before declaring the set of vulnerable releases in their vulnerability report. 
\minor{However, since many vulnerabilities were not analyzed by Snyk but only copied from other security trackers, we decided to manually inspect 50 additional vulnerability reports (25 from \npm and 25 from \rubygems), randomly chosen from the whole set of vulnerability reports that do not have a lower bound on the range of affected package releases. For each vulnerability, we manually checked its fixing commit and then verified whether the vulnerability was present in the first initial release~\cite{ozment2006milk}. Among these 50 cases, we found 8 false positives (4 for each ecosystem), \ie vulnerability reports in which the first initial version is claimed to be vulnerable, while it is not. This corresponds to a confidence interval of $0.81574 \pm 0.10356$ using Agresti-Coull's method~\cite{agresti1998approximate} with a confidence level of 95\%. This manual verification implies that our results overestimate the actual number of packages and package releases affected by reported vulnerabilities.}

\changed{
To construct our dataset of vulnerabilities to study in $RQ_{1,2}$, we have removed vulnerabilities affecting inactive packages and vulnerabilities that have been disclosed before 2017-04-17. Considering all vulnerabilities of all packages (\ie active and inactive) in the analysis might produce different outcomes.}

As a threat to \emph{internal validity}, when studying external \github projects that are not distributed via the package managers, we only focused on those \github projects explicitly mentioned in the \librariesio dataset. These projects were the most popular ones in terms of number of stars. While one may argue that this sample is not representative of all external projects depending on \npm or \rubygems packages, the selected projects do have $90\%$ of the total number of stars attributed to all possible project candidates available in the \librariesio dataset. We therefore consider the chosen set of \github projects to be representative for most \github projects that depend on \npm and \rubygems packages.

As a threat to \emph{conclusion validity}, we used metadata to identify vulnerable dependencies. This identification approach assumes that the metadata associated with the used dependencies (\eg package, version) and vulnerability descriptions (\eg affected package, list of affected versions) are always accurate. These metadata are used to map each package onto a list of known vulnerabilities that affect it. However, dependents that rely on a vulnerable package might only access functionality of the dependent package that is not affected by the vulnerability. Therefore, our results present an overestimation of the actual risk. Still, we believe \minor{in the importance of signalling maintainers of exposed dependents that they are relying on vulnerable dependencies.} It will be the responsibility of those maintainers to decide whether they are actually accessing vulnerable code, and to get rid of the vulnerable dependency if this happens to be the case.

As a threat to \emph{external validity}, our findings do not generalise to other package distributions (\eg Maven Central, CRAN, Cargo, PyPI). However, the design of our study can easily be replicated for other package distributions that are known to recommend \semver practices.

\changed{Another threat to \emph{external validity} stems from the fact that we relied on the manifests \emph{Gemfile} and \emph{packages.json} instead of the lockfiles \emph{Gemfile.lock} and \emph{package-lock.json} when extracting dependencies used in \github projects. The former are the default manifests that are always present in an \npm or \rubygems project and in which dependencies with their permissive or restrictive constraints are declared, while lockfiles list the specific releases of all dependencies that should be selected to replicate the deployment of the project. We do not know whether our results would remain the same when considering the dependencies specified in lockfiles rather than the ones specified in the manifest of each package. A manifest expresses (through dependency constraints) the set of releases that could be selected when the package is installed through the package manager. The latter always select the highest available release satisfying the constraints. As such, it may be the case that the specific releases pinned in a lockfile do not correspond to the ones that will be selected by the package manager, potentially leading to a different exposure to vulnerabilities. Although the use of a lockfile allows maintainers to explicitly select a non-vulnerable release of a package that they know to be vulnerable, the lockfile prevents from automatically benefiting from a fix in case the selected, pinned release is vulnerable. On the other hand, nothing prevents the maintainer from excluding vulnerable releases through appropriate dependency constraints in the package manifest, while still allowing future security patches to be selected as soon as they become available.}

\section{Conclusion}
\label{sec:conclusion}

This paper quantitatively analysed and compared how security vulnerabilities are treated in  \npm and \rubygems, 
two popular package distributions known to recommend the practice of semantic versioning.
Relying on the Snyk vulnerability database, we studied 2,786 vulnerabilities which affect 1,993 packages directly.
We observed that the number of reported vulnerabilities is increasing exponentially in \npm, and linearly in \rubygems.
Most of the reported vulnerabilities were of medium or high severity.
We observed that malicious packages occur much more frequently in \npm.%

\changed{
We analyzed the time needed to disclose and fix vulnerabilities and we found that half of the studied vulnerabilities needed more than two years to be disclosed, and more than four years to be fixed.}
Maintainers of reusable packages should therefore invest more effort in inspecting their packages for undiscovered vulnerabilities, especially if those packages have a lot of dependents that may get exposed to these vulnerabilities directly or indirectly.
Better tooling should be developed to help developers in looking for undiscovered vulnerabilities and fixing them.

By analysing the impact of vulnerable packages on dependent packages and dependent external projects (hosted on \github),
we observed that vulnerabilities can be found deep in the packages' and projects' dependency trees. 
Around one out of three packages and two out of three external projects are exposed to vulnerabilities coming from indirect dependencies. The most prevalent vulnerability type that affects \npm dependencies is \emph{Prototype Pollution}, while for \rubygems dependencies it is \emph{Denial of Service}. 

An important observation is that most of the vulnerabilities affecting dependencies of packages and external projects have known fixes (in the form of more recent package releases that are no longer vulnerable). Maintainers of dependents \changed{could} therefore invest more effort in checking their exposure to vulnerable dependencies, and in updating their outdated dependencies in order to reduce the number of dependency vulnerabilities. 
In contrast, the majority of the outdated and vulnerable dependencies need to be updated to a new major release to avoid the vulnerabilities.
We found that around one out of three vulnerabilities affecting direct dependencies of packages and external \github projects, \changed{might} be avoided by only making backward compatible dependency updates. 

\section*{Acknowledgments}

This research was partially funded by the Excellence of Science project 30446992 SECO-Assist financed by F.R.S.-FNRS and FWO-Vlaanderen, as well as FNRS Research Credit J015120 and FNRS Research Project T001718.
We express our gratitude to the security team of \emph{Snyk} for granting us permission to use their dataset of vulnerability reports for research purposes.

\balance
\providecommand{\noopsort}[1]{}

\bibliographystyle{unsrt}
\end{document}